\documentclass[onecolumn,11pt]{IEEEtran}
\usepackage{amssymb}
\usepackage{amsmath}
\usepackage{cases}
\usepackage{pslatex}
\usepackage{amsfonts,color,morefloats}
\usepackage{amssymb,amsmath,latexsym,amsthm}
\usepackage{hyperref}
\usepackage{multirow}
\usepackage{makecell}
\usepackage{amsfonts}
\usepackage{tikz}
\usetikzlibrary{mindmap}
\usetikzlibrary{chains}
\usepackage{indentfirst}
\usepackage{amsmath}
\usepackage{amsthm}
\usepackage{array}
\usepackage{amssymb}
\usepackage{color}
\usepackage{graphicx}
\usepackage[para]{threeparttable}
\usepackage{geometry}
\usepackage{verbatim}
\usepackage{framed}
\usepackage{bm}
\usepackage{arydshln}
\usepackage[mathscr]{eucal}
\usepackage{nicematrix}
\usepackage{dutchcal}

\newcommand{\wt}{{\mathrm{wt}}}

\newcommand{\tr}{{\mathrm{Tr}}}

\newcommand{\gf}{{\mathbb{F}}}

\newcommand{\C}{{\mathcal{C}}}

\newcommand{\vv}{\mathbf v}

\newcommand{\ee}{\mathbf e}
\newcommand{\cc}{\mathbf c}

\newcommand{\vct}[1]{\mathbf{#1}}
\newcommand{\R}{\mathcal R}
\newcommand{\Q}{\mathcal Q}

\DeclareMathOperator{\N}{N}

\newtheorem{theorem}{Theorem}
\newtheorem{lemma}{Lemma}

\newtheorem{corollary}{Corollary}

\newtheorem{definition}{Definition}
\newtheorem{example}{Example}

\newtheorem{remark}{Remark}

\begin{document}

\title{Generalized Covering radii and Deep Holes of a Family of Antiprimitive Narrow-Sense BCH Codes in Even Characteristic}
\author{Ketong Ren, Bing Liu, Cuiling Fan, Haode Yan*
\thanks{Ketong Ren and Cuiling Fan are with the School of Mathematics, Southwest Jiaotong University, Chengdu, 611756, China (e-mail: rktswjtu@163.com; e-mail: fcl@swjtu.edu.cn).}
\thanks{Bing Liu is with School of Computer and Software Engineering, Xihua University, Chengdu, 610039, China (liubing@mail.xhu.edu.cn).}
\thanks{Haode Yan is with the School of Science, Harbin Institute of Technology, Shenzhen, 518055, PR China (Corresponding author) (e-mail: yanhd@hit.edu.cn).}
}

\maketitle

\begin{abstract}
    This paper determines the generalized covering radii, deep holes, and complete coset leader weight distribution of the antiprimitive narrow-sense BCH code of length \(q+1\) and designed distance \(3\) over \(\mathbb{F}_q\), where \(q=2^m\) and \(m\geq 2\). We establish the complete generalized covering-radius hierarchy: the covering radius is \(3\), whereas the \(t\)-th generalized covering radius is \(4\) for every \(t\geq 2\). Using the covering radius, we then derive necessary and sufficient conditions for a received word to be a deep hole by reducing the syndrome representations to conditions involving roots of an associated cubic polynomial. An explicit parametrization yields complete, nonredundant representatives of all deep-hole cosets. Enumerating these representatives gives closed-form expressions for the numbers of deep holes and determines the complete coset leader weight distribution, with distinct formulas according to the parity of \(m\).
\end{abstract}


\section{Introduction}
    Let \(q\) be a prime power and let \(\mathcal C\) be an \([n,k]\) linear code over \(\mathbb F_q\). For \(\mathbf v\in\mathbb F_q^n\), its Hamming weight \(\operatorname{wt}(\mathbf v)\) is the number of its nonzero coordinates, and the Hamming distance between two words \(\mathbf u,\mathbf v\in\mathbb F_q^n\) is $d(\mathbf u,\mathbf v)=\operatorname{wt}(\mathbf u-\mathbf v)$. The error distance of a received word \(\mathbf v\) to \(\mathcal C\) is
        \[ d(\mathbf v,\mathcal C)=\min_{\mathbf c\in\mathcal C}d(\mathbf v,\mathbf c)=\min_{\mathbf c\in\mathcal C}\operatorname{wt}(\mathbf v-\mathbf c). \]
    Computing the error distance is essentially equivalent to solving the maximal likelihood decoding problem. The covering radius of \(\mathcal C\) is the maximum error distance
        \[ \rho(\mathcal C)=\max_{\mathbf v\in\mathbb F_q^n}d(\mathbf v,\mathcal C). \]
    Equivalently, it is the least integer \(r\) such that the Hamming balls of radius \(r\) centered at the codewords cover \(\mathbb F_q^n\). The covering radius has a natural generalization in which several syndromes are generated simultaneously by a common set of parity-check columns. The generalized covering radii of a linear code is defined as follows.
    \begin{definition}[{\cite[Definition~1]{tCRIT21}}]\label{def:tCR}
        Let $\mathcal{\C}$ be an $[n,k]$ linear code over $\mathbb{F}_q$ with a full-rank parity-check matrix $H\in\mathbb{F}_q^{(n-k)\times n}$. For an integer $t\ge 1$, the $t$-th generalized covering radius $\rho_t(\mathcal{\C})$ is the smallest nonnegative integer $r$ such that, for every $s_1,\ldots,s_t\in\mathbb{F}_q^{n-k}$, there exists a common set $I\subseteq\{1,\ldots,n\}$ with $|I|\le r$ satisfying
            \[ \{s_1,\ldots,s_t\}\subseteq \langle H_I\rangle_{\mathbb{F}_q}.\]
        Equivalently,
        \begin{equation}\label{eq:tCR}
            \rho_t(\mathcal{C})=\max_{s_1,\ldots,s_t\in\mathbb{F}_q^{n-k}}\min_{\substack{I\subseteq\{1,\ldots,n\}\\\{s_1,\ldots,s_t\}\subseteq \langle H_I\rangle_{\mathbb{F}_q}}}|I|.
        \end{equation}
        Here $H_I$ is the submatrix consisting of the columns of $H$ indexed by $I$, and $\langle H_I\rangle_{\mathbb{F}_q}$ denotes their $\mathbb{F}_q$-linear span.
    \end{definition}
    In particular, when $t=1$, $\rho_1(\mathcal{C})$ is exactly the covering radius $\rho(\mathcal{C})$. A received word $\vct v$ whose error distance attains this value is called a deep hole of $\C$; that is, $d(\vct v,\C)=\rho(\C)$. Thus, deep holes are the words at the greatest possible distance from the code. Let $\vct v\in\gf_q^n$. The weight of the coset $\vct v + \C$ is defined by 
        $$\operatorname{wt}(\mathbf v+C)=\min_{\mathbf c\in \C}\operatorname{wt}(\mathbf v+\mathbf c)=d(\mathbf v,C).$$ 
    A coset leader is a choice of an element $\vct l\in\gf_q^n$ of minimal weight in its coset, i.e., $\operatorname{wt}(\vct l) = \operatorname{wt}(\vct l + \C)$. A coset may contain several leaders, all of which have the same weight. In minimum-distance decoding, a coset leader represents an error pattern of minimum weight compatible with the received word. Let $N_i$ be the number of cosets of $\C$ of weight $i$. Equivalently, after choosing one leader from each coset, \(N_i(\mathcal C)\) is the number of chosen leaders of weight \(i\). The coset leader weight distribution of \(\mathcal C\) is the sequence of these numbers, ordered by increasing coset weight. The coset leader weight enumerator is the polynomial with coefficients $N_i$, i.e., $\sum_{i=0}^{\rho(\C)}N_i(\C)Z^i.$
    In particular, the deep-hole cosets are precisely those with coset leader weight \(\rho(\mathcal C)\). For classifying cosets and deep holes, we say that two vectors $\vv$ and $\mathbf w$ are equivalent if there exist $\lambda\in\gf_q^*$ and $\cc\in\C$ such that $ \mathbf w=\lambda\vv+\cc$. Clearly, this equivalence relation preserves the error distance to $\C$ and hence the coset leader weight. In this paper, we investigate the covering radius, deep holes, and coset leader weight distribution of the antiprimitive narrow-sense BCH code \(\mathcal C_{(q,q+1,3,1)}\) of length \(q+1\) and designed distance \(3\), where \(q\ge4\).
    
    The covering radius is a fundamental geometric parameter of linear codes, measuring their space-covering capability and having applications to data compression and decoding \cite{HaradaO2004,OrozcoJ2024}. Generalized covering radii extend this perspective, which quantifies the trade-off between storage, latency, and access complexity in linear database querying, with connections to generalized Hamming weights \cite{tCRIT21,Wei1991}. Computing the covering radius of a general linear code is both NP-hard and co-NP-hard \cite{McLoughlin1984}, and exact determinations remain challenging even for well-structured codes \cite{ShiTO2022,ShiHO2023,ShiLHO2025}. Once the covering radius \(\rho\) is known, classifying all deep holes amounts to identifying the syndromes that cannot be expressed as linear combinations of at most \(\rho-1\) columns of a parity-check matrix \cite{FangXZ2025,Kaipa2017,ZhangWK2020,ZhangW2023}. Determining the complete coset leader weight distribution extends this extremal analysis by enumerating cosets according to their minimum Hamming weights. The resulting distribution describes the number of cosets at every distance from the code and determines the probability of correct maximum-likelihood decoding on a \(q\)-ary symmetric channel \cite{BlokhuisPS2022,EzaharMM1987}.

    Covering-radius research has developed along two complementary strands: the exact determination of covering radii for codes, and the derivation of general bounds. For the classical covering radius, exact values are known only for a few families, while most general results are bounds. Exact covering radii have been obtained for Melas codes \cite{ShiTO2022} and generalized Zetterberg-type codes \cite{ShiHO2023,ShiLHO2025,XiongY2026}. More recently, the generalized covering radius hierarchy introduced by Elimelech, Firer and Schwartz \cite{tCRIT21} has become an active research topic. Nevertheless, exact values and the full generalized covering-radius hierarchy still remain open for many code families. Deep-hole research mainly focuses on MDS codes and its extensions. For RS codes, Kaipa linked deep-hole classes to MDS extensions of dual GRS codes, proving the Cheng--Murray conjecture for $k\ge \lfloor (q-1)/2\rfloor$ \cite{Kaipa2017}. For PRS codes, Zhang et al. constructed three classes via automorphisms and normal rational curve geometry, and proved the completeness \cite{ZhangWK2020}. Generalizations include twisted RS codes, with covering radius and standard deep holes determined and completeness proved in ranges \cite{FangXZ2025}; Gabidulin codes with tight bounds \cite{FangWW2018}; and elliptic curve codes, with explicit constructions and a completeness conjecture \cite{ZhangW2023}. Coset leader weight enumerators further refine these problems: Helleseth gave general extension-code results \cite{Helleseth1979}; Hou classified $\mathrm{RM}(m-3,m)$ cosets and gave recursive weight formulas \cite{Hou1994}; and the extended enumerator for the twisted-cubic GRS $[q+1,q-3,5]_q$ code was computed \cite{BlokhuisPS2022}.
    
    BCH codes are of particular interest in this context because they combine prescribed minimum-distance guarantees with efficient algebraic encoding and decoding for communication and storage \cite{DingLi2024,HuffmanPless2003}. For binary primitive narrow-sense double- and triple-error-correcting BCH codes of length $2^m-1$ with $m\ge5$, the covering radii are $3$ and $5$, respectively \cite{GorensteinEtAl1960,Helleseth1978}; their complete coset-leader weight distributions of both the original and extended codes are known \cite{Charpin1994,CharpinHellesethZinoviev2006}. More generally, for each fixed $t\ge2$, the binary primitive narrow-sense BCH code of designed distance $2t+1$ has covering radius $2t-1$ for all sufficiently large $m$ \cite{Cohen1997}, with improved sufficient bounds on $m$ obtained in \cite{KavutTutdere2019}. Related multicovering and $2$-covering radii have also been investigated, and the $2$-covering radius of $\mathrm{BCH}(2,m)$ has been determined for sufficiently large $m$ \cite{Klapper2006}. Within the generalized covering-radius hierarchy of Elimelech, Firer, and Schwartz \cite{tCRIT21}, the second and third generalized covering radii of $\mathrm{BCH}(2,m)$ were determined \cite{Ozbudak2026,Yohananov2025}, and a generalized supercode lemma was used to simplify lower-bound proofs and to establish the asymptotic range $2k\le \rho_k(\mathrm{BCH}(2,m))\le 2k+1$ \cite{XiongCHY2026}. Beyond primitive binary BCH codes, antiprimitive BCH codes of length $q^s+1$ over $\mathbb{F}_q$ have a reversible linear complementary dual structure \cite{LiDingLi2017}, and dimensions and minimum distances yielding families with good or optimal parameters have been determined \cite{LiuDingLi2017,ZhuEtAl2022}; certain families of length $q+1$ admit $3$-transitive automorphism groups and support Witt spherical geometry designs \cite{LiuEtAl2022}. For the designed-distance-three code $\mathcal{C}_{(q,q+1,3,1)}$, minimum-distance properties have been investigated in detail \cite{XuEtAl2025a}; moreover, for $q=2^m$ with even $m\ge4$, this code is near maximum distance separable and has connections to optimal locally recoverable codes \cite{DingTang2020,TanEtAl2023}.

    These parameter and structural results motivate a complementary study of distances from arbitrary received words and of the associated coset structure of $\mathcal{C}_{(q,q+1,3,1)}$. In this paper, we prove that its covering radius is \(3\), whereas its \(t\)-th generalized covering radius is \(4\) for every \(t\ge2\). We characterize all deep-hole syndromes and construct complete, nonredundant representatives for coset classes. Their explicit enumeration yields the complete coset-leader weight distribution, with closed-form expressions depending on the parity of \(m\).
    
    This paper is organized as follows. Section~\ref{sec:Pre} introduces the notation and preliminaries and constructs explicit coset representatives. Section~\ref{sec:CR} determines the covering radius and all generalized covering radii of the code. Section~\ref{sec:DH} characterizes all deep holes and provides nonredundant representatives and enumeration formulas for the deep-hole cosets. Section~\ref{sec:CLWD} derives the complete coset-leader weight distribution. Finally, Section~\ref{sec:con} concludes the paper.

\section{Preliminary}\label{sec:Pre}
Throughout this paper, let \(q=2^m\) with \(m\ge2\), and let \(\mathbb F_q\) and \(\mathbb F_{q^2}\) denote the finite fields of orders \(q\) and \(q^2\), respectively. For \(z\in\mathbb F_{q^2}\), let $\bar z$ denote $z^q$ and the norm function from $\gf_{q^2}$ to $\gf_q$ is denoted by
    \[ \N(z)=\N_{\mathbb F_{q^2}/\mathbb F_q}(z)=z\bar z. \]
Let \(\operatorname{Tr}=\operatorname{Tr}_{\mathbb F_q/\mathbb F_2}\) denote the absolute trace from $\gf_q$ to $\gf_2$, given by
    \[ \operatorname{Tr}(\alpha)=\sum_{i=0}^{m-1}\alpha^{2^i}, \text{ where } \alpha\in\mathbb F_q. \]
Let
    \[ U_{q+1}=\{x\in\mathbb F_{q^2}^{*}:x^{q+1}=1\}, \]
and fix a generator \(u\) of this cyclic group of order \(q+1\). In particular, \(\bar x=x^{-1}\) for every \(x\in U_{q+1}\). 

\subsection{BCH Codes}
    Let \(n\ge2\) satisfy \(\gcd(n,q)=1\). Under the polynomial representation, every cyclic code \(\mathcal C\) of length \(n\) over \(\mathbb F_q\) has the form
        \[ \C=\langle g(X)\rangle\subseteq \mathbb F_q[X]/(X^n-1), \]
    where \(g(X)\) is its unique monic generator polynomial dividing\(X^n-1\). Choose a primitive \(n\)-th root of unity \(\mathbcal{e}\in\mathbb F_{q^{\operatorname{ord}_n(q)}}\). For each integer \(i\), denote by \(M_{\mathbcal{e}^i}(X)\) the minimal polynomial of \(\mathbcal{e}^i\) over \(\mathbb F_q\).
    
    For \(2\le\delta\le n\), the narrow-sense BCH code \(\mathcal C_{(q,n,\delta,1)}\) is the cyclic code with generator polynomial
        \[ g_{(q,n,\delta,1)}(X)=\operatorname{lcm}\bigl(M_{\mathbcal{e}}(X),M_{\mathbcal{e}^2}(X),\ldots,M_{\mathbcal{e}^{\delta-1}}(X) \bigr), \]
    where the least common multiple is taken in \(\mathbb F_q[X]\).
    
    In this paper, we take \(n=q+1\), \(\delta=3\), and \(\mathbcal{e}=u\), where \(u\) is the generator of \(U_{q+1}\) fixed above. Thus, \(\mathcal C_{(q,q+1,3,1)}\) is referred to as a class of antiprimitive narrow sense BCH code \cite{Ding2016}. Since \(u^q=u^{-1}\) and the four elements \(u,u^{-1},u^2,u^{-2}\) are distinct for \(q\ge4\), \(M_u(X)\) and \(M_{u^2}(X)\) are distinct irreducible quadratic polynomials. Consequently,
        \[ g_{(q,q+1,3,1)}(X)=\operatorname{lcm}\bigl(M_u(X),M_{u^2}(X)\bigr)=M_u(X)M_{u^2}(X). \]
    The generator polynomial therefore has degree four, giving $\dim_{\mathbb F_q}\mathcal C_{(q,q+1,3,1)}=q-3.$

    With the coordinate order fixed above, an extension-field parity-check matrix of \(\mathcal C_{(q,q+1,3,1)}\) is
    \[ H_{(q,q+1,3,1)}=
        \begin{pmatrix}
            1 & u   & u^2 & \cdots & u^q\\
            1 & u^2 & u^4 & \cdots & u^{2q}
        \end{pmatrix}. \]
    Equivalently,
    \[ \mathcal C_{(q,q+1,3,1)}=
        \left\{
            \mathbf c\in\mathbb F_q^{q+1}:
            H_{(q,q+1,3,1)}\mathbf c^{\mathsf T}=0
        \right\}.\]
    Expanding the two syndrome coordinates with respect to an \(\mathbb F_q\)-basis of \(\mathbb F_{q^2}\) yields a \(4\times(q+1)\) parity-check matrix of rank four over \(\mathbb F_q\).

\subsection{Auxiliary Results on the multiplicative subgroup}
We collect auxiliary results on the multiplicative subgroup \(U_{q+1}\) that will be used to determine the covering radius and characterize deep holes. The following results are useful in determining the solvability of a system concerning proving the desired upper bound of \(\rho(\C_{(q,q+1,3,1)})\).
\begin{lemma}\label{lem:U_to_F}\cite{Tzr2018}
    Let $ a \in \gf_{q^{2}}\setminus\gf_{q}$ be fixed. Then
        $$U_{q+1}\setminus\{1\}=\left\{\frac{\alpha+\overline{a}}{\alpha+a} : \alpha\in\gf_{q}\right\}.$$
\end{lemma}

Along with this parametrization, we will use the following criterion for determining the number of roots of a quadratic equation in \(U_{q+1}\).
\begin{lemma}\label{lem:sol_in_unit_peq2}\cite{Tzr2018}
	Let $a,b\in \gf^*_{q^2}$ satisfying $\tr_{\gf_{q^2}/\gf_2}\left(\frac{b}{a^2}\right)=0$. Then the quadratic equation $x^2+ax+b=0$ has
	    \begin{itemize}
		    \item both solutions in $U_{q+1}$ if and only if
			    $$b=\frac{a}{\overline{a}},~\tr\left(\frac{b}{a^2}\right)=\tr\left(\frac{1}{a\overline{a}}\right)=1.$$
		    \item exactly one solution in $U_{q+1}$ if and only if
			    $$b\neq \frac{a}{\overline{a}},~(1+b\overline{b})(1+a\overline{a}+b\overline{b})+a^2\overline{b}+\overline{a}^2b=0.$$
	    \end{itemize}
\end{lemma}

As a consequence of Lemma \ref{lem:sol_in_unit_peq2}, we obtain the following identity on \(U_{q+1}\).
\begin{lemma} \label{lem:values_for_addsum}
    When $q$ is even, if $u\in U_{q+1}\setminus\{1\}$, then it holds that $\tr\left(\frac{1}{u+u^{-1}}\right)=1.$
\end{lemma}
\begin{proof}
   Since $u$ and $u^{-1}$ are two distinct solutions in $U_{q+1}$ of $x^2+(u+u^{-1})x+1=0$, the necessary and sufficient condition in Lemma \ref{lem:sol_in_unit_peq2} immediately yields $\tr\bigl(1/(u+u^{-1})^2\bigr)=\tr\bigl(1/(u+u^{-1})\bigr)=1$.
\end{proof}

\subsection{Covering Radii and Coset Representatives}

    We first recall the parity-check characterization of the covering radius and the basic properties of generalized covering radii.
    \begin{lemma}\label{lem:CR_PCmatrix}\cite[Theorem 11.1.2]{HuffmanPless2003}
        If $\C$ is an $[n,k]$ linear code over $\gf_q$ with parity check matrix $H$, then $\rho(\C)$ is the smallest number such that every non-zero vector in $\gf_q^{n-k}$ is a combination of $\rho(\C)$ or fewer columns of $H$.
    \end{lemma}
    
    \begin{lemma}\cite{tCRIT21}\label{lem:t_CR_PCmatrix}
        For any $t\in\mathbb{Z}^{+}$, the $t$-th generalized covering radius of a linear code $\mathcal{\C}$ satisfies:
        $$\rho_1(\mathcal{\C})\leqslant \rho_2(\mathcal{\C})\leqslant \cdots \leqslant \rho_{n-k}(\mathcal{\C})=\rho_{s}(\mathcal{C})=n-k, \quad s\in\mathbb{Z}^{+},\ s>n-k.$$
    \end{lemma}
    For the code $\C_{(q,q+1,3,1)}$, the redundancy is $(q+1)-(q-3)=4$. Hence Lemma~\ref{lem:t_CR_PCmatrix} gives
        \[ \rho_t\bigl(\C_{(q,q+1,3,1)}\bigr)\leqslant4 \]
    for all $t\geqslant1$.

    We next describe \(\C_{(q,q+1,3,1)}\)'s syndromes and construct explicit coset representatives. For $\mathbf v=(v_0,\ldots,v_q)\in\gf_q^{q+1}$, let $v(X)=\sum_{j=0}^{q}v_jX^j.$ Then the syndrome of \(\vct v\) with respect to \(\C_{(q,q+1,3,1)}\) is
        \[ \operatorname{Syn}(\mathbf v)=H_{(q,q+1,3,1)}\mathbf v^{\mathsf T}=\bigl(v(u),v(u^2)\bigr). \]
    For any $\mathbf v,\mathbf e\in\gf_q^{q+1}$, $\operatorname{Syn}(\mathbf e)=\operatorname{Syn}(\mathbf v)$ if and only if $\mathbf v-\mathbf e\in\C_{(q,q+1,3,1)}$. It follows that, for every $\mathbf v\in\gf_q^{q+1}$,
    \begin{equation}\label{eq:distance}
        d(\mathbf v,\C_{(q,q+1,3,1)})=\min_{\substack{\mathbf e\in\gf_q^{q+1}\\\operatorname{Syn}(\mathbf e)=\operatorname{Syn}(\mathbf v)}}\wt(\mathbf e).
    \end{equation}
    To construct explicit representatives of the syndromes, we use Lagrange interpolation on $\Omega=\{u,u^{-1},u^2,u^{-2}\}$. Since $u$ has order $q+1\geqslant5$, these four elements are distinct. Define
    \begin{align}\label{eq:lagrange-basis}
        L_u(X)=\frac{1}{(u-u^{-1})(u-u^2)(u-u^{-2})}(X-u^{-1})(X-u^2)(X-u^{-2}).
    \end{align}
    $L_{u^{-1}}$, $L_{u^2}$ and $L_{u^{-2}}$ are defined similarly. For $(a,b)\in \gf_{q^2}^2$, let
    \begin{equation}\label{eq:interpolation-polynomial}
        F_{a,b}(X)=aL_u(X)+a^qL_{u^{-1}}(X)+bL_{u^2}(X)+b^qL_{u^{-2}}(X).
    \end{equation}
    Let $F_{a,b}(X)=\sum_{j=0}^{3}f_j(a,b)X^j$, and put
    \begin{equation}\label{eq:representative}
        \vct e(a,b)=\bigl(f_0(a,b),f_1(a,b),f_2(a,b),f_3(a,b),0,\ldots,0\bigr),
    \end{equation}
    where the last $q-3$ coordinates are zero. Formula~\eqref{eq:representative} gives explicit coset representatives in the original BCH coordinates.

    \begin{lemma}\label{lem:interpolation}
    For every $(a,b)\in\gf_{q^2}^{2}$, the polynomial $F_{a,b}$ belongs to $\gf_q[X]$ and has degree at most three. Moreover, $\mathbf e(a,b)\in\gf_q^{q+1}$ and
        \[ \operatorname{Syn}\bigl(\mathbf e(a,b)\bigr)=(a,b). \]
    Consequently, every syndrome has a representative supported on coordinates $0,1,2,3$.
    \end{lemma}
    \begin{proof}
        It is easy to check that $\overline{L_{u}(X)}=L_{u^{-1}}(X)$. Then for every $z\in\Omega$ we have $\overline{L_z}(X)=L_{\bar z}(X)$. The $q$-th power of $F_{a,b}$ therefore interchanges the terms defining $F_{a,b}$ in pairs, giving $\overline{F_{a,b}}=F_{a,b}$. Thus all coefficients of $F_{a,b}$ lie in $\gf_q$. Since each $L_z$ has degree three, $\deg F_{a,b}\leqslant3$. By Lagrange interpolation, $L_z(z)=1$ and $L_z(h)=0$ for distinct $z,h\in\Omega$. Hence
            \[ F_{a,b}(u)=a,\qquad F_{a,b}(u^2)=b, \]
        and consequently
            \[ \operatorname{Syn}\bigl(\mathbf e(a,b)\bigr)=\bigl(F_{a,b}(u),F_{a,b}(u^2)\bigr)=(a,b). \]
        The support assertion follows from the definition of $\mathbf e(a,b)$.
    \end{proof}
    
    For $0\leqslant i\leqslant\rho(\C_{(q,q+1,3,1)})$, the correspondence between syndromes and cosets yields
        \[ N_i(\C_{(q,q+1,3,1)})=\#\bigl\{(a,b)\in\mathbb F_{q^2}^{\,2}:d(\vct{e}(a,b),\C_{(q,q+1,3,1)})=i\bigr\}. \]
    Since error distance is constant on each coset and each coset contains $q^{q-3}$ vectors, we also have
        \[ \#\{\vct{v}\in\mathbb F_q^n:d(\vct{v},\C_{(q,q+1,3,1)})=i\}=q^{q-3}N_i(\C_{(q,q+1,3,1)}). \]

\section{Generalized Covering Radius}\label{sec:CR}
In this section, we determine the complete generalized covering-radius hierarchy of \(\mathcal C_{(q,q+1,3,1)}\) for \(q=2^m\ge4\). We first establish that its covering radius is \(3\). We then prove that its \(t\)-th generalized covering radius is \(4\) for every \(t\ge2\).

\subsection{Covering Radius}
In this subsection, we determine the covering radius of \(\mathcal{C}_{(q,q+1,3,1)}\) when \(q\) is even. We first give the tight lower bound.
\begin{lemma}\label{lem:rho_lower_bound}
    Let $q\ge4$ be a prime power. It holds that
            \[\rho(\C_{(q,q+1,3,1)})\ge3.\]
\end{lemma}
\begin{proof}
    Since $\dim_{\gf_q}\C_{(q,q+1,3,1)}=q-3$, the number of cosets of $\C_{(q,q+1,3,1)}$ in $\gf_q^{q+1}$ is $q^4$. On the other hand, the number of vectors in $\gf_q^{q+1}$ of Hamming weight at most two is
        \[ 1+(q+1)(q-1)+\binom{q+1}{2}(q-1)^2=q^4-\frac{q(q-1)(q+1)^2}{2}<q^4. \]
    Consequently, fewer than $q^4$ cosets contain a vector of weight at most two. Thus at least one coset has minimum weight at least three, and hence $\rho\bigl(\C_{(q,q+1,3,1)}\bigr)\geqslant3$.
\end{proof}

In view of the lower bound in Lemma \ref{lem:rho_lower_bound}, it remains to establish an upper bound of three, which is dealt with in the following system.
\begin{lemma}\label{lem:qeven_rho_ubound}
Assume that $q>4$. For any \(b_1\in\gf_q, b_2 \in \gf_{q^2}\), there exist \(\alpha_1, \alpha_2, \alpha_3 \in \gf_q\) and \(u_1, u_2, u_3 \in U_{q+1}\) such that the following system of equations holds:
\begin{equation}\label{eqtsys:peq2_heq1_1}
    \begin{cases}
        \alpha_1u_1+\alpha_2u_2+\alpha_3u_3 = b_1, \\
        \alpha_1^2u_1+\alpha_2^2u_2+\alpha_3^2u_3 = b_2.
    \end{cases}
\end{equation}
\end{lemma}
\begin{proof}
    If $b_2=b_1^2$, then the assertion is immediate by taking $\alpha_1=b_1, \alpha_2=\alpha_3=0, u_1=1, u_2,u_3\in U_{q+1}$. In the following, assume that $b_2\neq b_1^2$.
    The proof proceeds by setting $u_3=1$. Then it is enough to find $\alpha_1,\alpha_2,\alpha_3\in F_q$ and distinct elements $u_1,u_2\in U_{q+1}\setminus\{1\}$ satisfying 
    \begin{equation}\label{eqtsys:peq2_heq1_2}
        \begin{cases} 
            \alpha_1u_1+\alpha_2u_2+\alpha_3=b_1,\\ 
            \alpha_1^2u_1+\alpha_2^2u_2+\alpha_3^2=b_2. 
        \end{cases}
    \end{equation}
    From the first equation in \eqref{eqtsys:peq2_heq1_2} we obtain
        $$\begin{cases} 
            \alpha_1u_1+\alpha_2u_2=b_1+\alpha_3,\\ 
            \alpha_1u_1^{-1}+\alpha_2u_2^{-1}=b_1+\alpha_3. 
        \end{cases}$$
    Since $u_1\neq u_2$, applying Cramer's rule yields
    \begin{equation}\label{eqtsys:peq2_heq1_3}
	    \begin{cases}
		    \alpha_1=\frac{u_2^{-1}(b_1+\alpha_3)+u_2(b_1+\alpha_3)}{u_1^{-1}u_2+u_1u_2^{-1}} \\
		    \alpha_2=\frac{u_1^{-1}(b_1+\alpha_3)+u_1(b_1+\alpha_3)}{u_1^{-1}u_2+u_1u_2^{-1}}.
	    \end{cases}
    \end{equation}
    Since the two expressions are fixed by the $q$-th power map, we have $\alpha_1,\alpha_2\in F_q$.
    Similarly, from the second equation in \eqref{eqtsys:peq2_heq1_1} we obtain the following system:
    \begin{equation}\label{eqtsys:peq2_heq1_4}
	    \begin{cases}
		    \alpha_1^2u_1+\alpha_2^2u_2= b_2+\alpha_3^2 \\
		    \alpha_1^2u_1^{-1}+\alpha_2^2u_2^{-1}= \overline{b_2}+\alpha_3^2.
	    \end{cases}
    \end{equation}
    Again by Cramer's rule,
    \begin{equation}\label{eqtsys:peq2_heq1_5}
	    \begin{cases}
		    \alpha_1^2=\frac{u_2^{-1}(b_2+\alpha_3^2)+u_2(\overline{b_2}+\alpha_3^2)}{u_1^{-1}u_2+u_1u_2^{-1}} \\
		    \alpha_2^2=\frac{u_1^{-1}(b_2+\alpha_3^2)+u_1(\overline{b_2}+\alpha_3^2)}{u_1^{-1}u_2+u_1u_2^{-1}}.
	    \end{cases}
    \end{equation}
    From (\ref{eqtsys:peq2_heq1_3}) and (\ref{eqtsys:peq2_heq1_5}) we deduce two equations on $\alpha_3^2$:
    \begin{equation}\label{eqtsys:peq2_heq1_a_3}
        \begin{cases}
            \frac{(u_1+1)(u_2+1)^2(u_2^2+u_1)}{u_1u_2^2}\alpha_3^2=(u_2^{-2}+u_2^2)b_1^2+u_1^{-1}b_2+u_1\overline{b_2}+u_1u_2^{-2}b_2+u_1^{-1}u_2^2\overline{b_2}.\\
            \frac{(u_1+1)^2(u_2+1)(u_1^2+u_2)}{u_1^2u_2}\alpha_3^2=(u_1^{-2}+u_1^2)b_1^2+u_2^{-1}b_2+u_2\overline{b_2}+u_1^{-2}u_2b_2+u_1^{2}u_2^{-1}\overline{b_2}.
        \end{cases}
    \end{equation}
    Combining these two equations, we have
    \begin{align*}
        &\frac{(u_1+1)(u_1^2+u_2)}{u_1}((u_2^{-2}+u_2^2)b_1^2+u_1^{-1}b_2+u_1\overline{b_2}+u_1u_2^{-2}b_2+u_1^{-1}u_2^2\overline{b_2})\\
        &+\frac{(u_2+1)(u_2^2+u_1)}{u_2}((u_1^{-2}+u_1^2)b_1^2+u_2^{-1}b_2+u_2\overline{b_2}+u_1^{-2}u_2b_2+u_1^{2}u_2^{-1}\overline{b_2})=0.
    \end{align*}
    That is,
    \begin{equation}\label{eqt:peq2_heq1_1}
        \begin{aligned}
            &(u_1+u_2+1+u_1u_2(u_1u_2+u_1+u_2))b_1^2+(u_1^2+u_2^2+1+u_1+u_2+u_1u_2)b_2\\
            &+(u_1^2+u_2^2+u_1^2u_2^2+u_1^2u_2+u_1u_2^2+u_1u_2)\overline{b_2}=0.
        \end{aligned}
    \end{equation}
    Note that $u_2=1$ satisfies \eqref{eqt:peq2_heq1_1} only if $(b_2+\overline{b_2})(u_1+1)^2=0$ and $u_2=u_1$ satisfies \eqref{eqt:peq2_heq1_1} only if $u_1$ satisfies
        $$(b_1^2+\overline{b_2})u_1^4+(b_2+\overline{b_2})u_1^2+b_1^2+b_2=0.$$
    If $(b_1^2+\overline{b_2})u_1^2+(b_1^2+\overline{b_2})u_1+b_2+\overline{b_2}\neq 0$, then \eqref{eqt:peq2_heq1_1} reduces to
    \begin{equation}\label{eqt:peq2_heq1_3}
        u_2^2 + \frac{(b_1^2+\overline{b_2})u_1^2+(b_2+\overline{b_2})u_1+b_1^2+b_2}{(b_1^2+\overline{b_2})u_1^2+(b_1^2+\overline{b_2})u_1+b_2+\overline{b_2}}u_2+ \frac{(b_2+\overline{b_2})u_1^2+(b_2+b_1^2)u_1+b_1^2+b_2}{(b_1^2+\overline{b_2})u_1^2+(b_1^2+\overline{b_2})u_1+b_2+\overline{b_2}}=0.
    \end{equation}
    Set $e_1=\frac{(b_1^2+\overline{b_2})u_1^2+(b_2+\overline{b_2})u_1+b_1^2+b_2}{(b_1^2+\overline{b_2})u_1^2+(b_1^2+\overline{b_2})u_1+b_2+\overline{b_2}}$, $e_0=\frac{(b_2+\overline{b_2})u_1^2+(b_2+b_1^2)u_1+b_1^2+b_2}{(b_1^2+\overline{b_2})u_1^2+(b_1^2+\overline{b_2})u_1+b_2+\overline{b_2}}$. One may verify that 
        $$\frac{e_0}{e_1^2}\in\gf_q, \text{ and } e_0 = \frac{e_1}{\overline{e_1}}.$$
    Consequently, by Lemma \ref{lem:sol_in_unit_peq2}, \eqref{eqt:peq2_heq1_3} has solutions in $U_{q+1}$ if and only if there exist $u_1\in U_{q+1}\setminus \{1\}$ satisfying $(b_1^2+\overline{b_2})u_1^4+(b_2+\overline{b_2})u_1^2+b_1^2+b_2\neq0$ and $(b_1^2+\overline{b_2})u_1^2+(b_1^2+\overline{b_2})u_1+b_2+\overline{b_2}\neq0$ such that 
    \begin{equation}\label{eqt:peq2_heq1_4}
        \tr\left(\frac{e_0}{e_1^2}\right)=\tr\left(\frac{[(b_2+\overline{b_2})u_1^2+(b_1^2+b_2)u_1+(b_1^2+b_2)][(b_1^2+\overline{b_2})u_1^2+(b_1^2+\overline{b_2})u_1+b_2+\overline{b_2}]}{[(b_1^2+\overline{b_2})u_1^2+(b_2+\overline{b_2})u_1+b_1^2+b_2]^2}\right)=1.
    \end{equation}
    Set $R_1=\{u \in U_{q+1} : (b_1^2+\overline{b_2})u^2+(b_2+\overline{b_2})u+b_1^2+b_2=0\}$ and $R_2=\{u \in U_{q+1} : (b_1^2+\overline{b_2})u^4+(b_2+\overline{b_2})u^2+b_1^2+b_2=0\}$. Then 
    $$\begin{cases}
        R_1\cap R_2=\{1\},\\
        R_1=R_2=U_{q+1}, \text{ when } b_2=b_1^2,\\
        R_1=R_2=\{1\}, \text{ when } b_2\in\gf_q\setminus\{b_1^2\},\\
        |R_2\setminus R_1|=1,\text{ when } b_2\in\gf_{q^2}\setminus\gf_q.
    \end{cases}$$
    The last equation holds due to $\tr\left(\frac{(b_1^2+\overline{b_2})(b_1^2+b_2)}{(\overline{b_2}+b_2)^2}\right)=\tr^s_1\left(\frac{b_2\overline{b_2}}{(\overline{b_2}+b_2)^2}\right)=1.$
    \begin{itemize}
        \item When $b_2\in\gf_q\setminus\{b_1^2\}$, 
                \begin{align*}
                    &\sum_{u_1\in U_{q+1}\setminus (R_1\cup R_2)}(-1)^{\tr\left(\frac{[(b_2+\overline{b_2})u_1^2+(b_1^2+b_2)u_1+(b_1^2+b_2)][(b_1^2+\overline{b_2})u_1^2+(b_1^2+\overline{b_2})u_1+b_2+\overline{b_2}]}{[(b_1^2+\overline{b_2})u_1^2+(b_2+\overline{b_2})u_1+b_1^2+b_2]^2}\right)}\\
                    =&\sum_{u_1\in U_{q+1}\setminus \{1\}}(-1)^{\tr\left(\frac{1}{u_1+u_1^{-1}}\right)}\\
                    =&-q.
                \end{align*}
        \item When $b_2\in\gf_{q^2}\setminus\gf_q$, let $A$ be an arbitrary element in $\gf_{q^2}\setminus\gf_q$. 
                Note that when $u_1\in R_2\setminus \{1\}$, $(b_1^2+\overline{b_2})u_1^2+b_1^2+b_2=0$ since $(b_1^2+\overline{b_2})u_1^4+(b_2+\overline{b_2})u_1^2+b_1^2+b_2=((b_1^2+\overline{b_2})u_1^2+b_1^2+b_2)(u_1+1)^2$. Then 
                \begin{align*}
                    &\sum_{u_1\in R_2\setminus \{1\}}(-1)^{\tr\left(\frac{[(b_2+\overline{b_2})u_1^2+(b_1^2+b_2)u_1+(b_1^2+b_2)][(b_1^2+\overline{b_2})u_1^2+(b_1^2+\overline{b_2})u_1+b_2+\overline{b_2}]}{[(b_1^2+\overline{b_2})u_1^2+(b_2+\overline{b_2})u_1+b_1^2+b_2]^2}\right)}\\
                    =&\sum_{u_1\in R_2\setminus \{1\}}(-1)^{\tr\left(\frac{(b_1^2+\overline{b_2})^2u_1^2+(b_1^2+b_2)^2}{(b_2+\overline{b_2})^2u_1}\right)}\\
                    =&\sum_{u_1\in R_2\setminus \{1\}}(-1)^{\tr\left(\frac{(b_1^2+b_2)}{(b_2+\overline{b_2})u_1}\right)}\\
                    =&\sum_{u_1\in R_2\setminus \{1\}}(-1)^{\tr\left(\frac{(b_1^2+\overline{b_2})(b_1^2+b_2)}{(b_2+\overline{b_2})^2}\right)}\\
                    =&-1.
                \end{align*}
            Then by Lemma \ref{lem:U_to_F},
                \begin{align*}
                    &\sum_{u_1\in U_{q+1}\setminus (R_1\cup R_2)}(-1)^{\tr\left(\frac{[(b_2+\overline{b_2})u_1^2+(b_1^2+b_2)u_1+(b_1^2+b_2)][(b_1^2+\overline{b_2})u_1^2+(b_1^2+\overline{b_2})u_1+b_2+\overline{b_2}]}{[(b_1^2+\overline{b_2})u_1^2+(b_2+\overline{b_2})u_1+b_1^2+b_2]^2}\right)}\\
                    =&\sum_{u_1\in U_{q+1}\setminus R_1}(-1)^{\tr\left(\frac{[(b_2+\overline{b_2})u_1^2+(b_1^2+b_2)u_1+(b_1^2+b_2)][(b_1^2+\overline{b_2})u_1^2+(b_1^2+\overline{b_2})u_1+b_2+\overline{b_2}]}{[(b_1^2+\overline{b_2})u_1^2+(b_2+\overline{b_2})u_1+b_1^2+b_2]^2}\right)}\\
                    &-\sum_{u_1\in R_2\setminus \{1\}}(-1)^{\tr\left(\frac{[(b_2+\overline{b_2})u_1^2+(b_1^2+b_2)u_1+(b_1^2+b_2)][(b_1^2+\overline{b_2})u_1^2+(b_1^2+\overline{b_2})u_1+b_2+\overline{b_2}]}{[(b_1^2+\overline{b_2})u_1^2+(b_2+\overline{b_2})u_1+b_1^2+b_2]^2}\right)}\\
                    =&\sum_{x_1\in\gf_q\setminus \{\frac{(b_1^2+\overline{b_2})\overline{A}^2+(b_1^2+b_2)A^2+(b_2+\overline{b_2})A\overline{A}}{(b_2+\overline{b_2})(A+\overline{A})}\}}(-1)^{\tr\left(\frac{1}{(A+\overline{A})^2}x_1^2+\frac{1}{A+\overline{A}}x_1+1+\frac{A\overline{A}}{(A+\overline{A})^2}+\frac{M_1}{M_0}\right)}+1,
                \end{align*}
            where
                \begin{align*}
                    M_1&=(b_1^2+b_2)(b_1^2+\overline{b_2})(A+\overline{A})^4.\\
                    M_0&=[(b_2+\overline{b_2})(A+\overline{A})x_1+(b_1^2+\overline{b_2})\overline{A}^2+(b_1^2+b_2)A^2+(b_2+\overline{b_2})A\overline{A}]^2.
                \end{align*}
                Since $q$ is even, there exists a unique $\sqrt{M_1}$ such that $(\sqrt{M_1})^2=M_1$. then:
                \begin{itemize}
                    \item When $m$ is even,
                            \begin{align*}
                                &\sum_{x_1\in\gf_q\setminus \{\frac{(b_1^2+\overline{b_2})\overline{A}^2+(b_1^2+b_2)A^2+(b_2+\overline{b_2})A\overline{A}}{(b_2+\overline{b_2})(A+\overline{A})}\}}(-1)^{\tr\left(\frac{1}{(A+\overline{A})^2}x_1^2+\frac{1}{A+\overline{A}}x_1+1+\frac{A\overline{A}}{(A+\overline{A})^2}+\frac{M_1}{M_0}\right)}+1\\
                                =&(-1)\sum_{x_1\in\gf_q\setminus \{\frac{(b_1^2+\overline{b_2})\overline{A}^2+(b_1^2+b_2)A^2+(b_2+\overline{b_2})A\overline{A}}{(b_2+\overline{b_2})(A+\overline{A})}\}}(-1)^{\tr\left(\frac{\sqrt{M_1}}{(b_2+\overline{b_2})(A+\overline{A})x_1+(b_1^2+\overline{b_2})\overline{A}^2+(b_1^2+b_2)A^2+(b_2+\overline{b_2})A\overline{A}}\right)}+1\\
                                =&(-1)\sum_{t_1\in\gf_q^*}(-1)^{\tr\left(t_1\right)}+1\\
                                =&(-1)\left(\sum_{t_1\in\gf_q}(-1)^{\tr\left(t_1\right)}-1\right)+1\\
                                =&2.
                            \end{align*}
                    \item When $m$ is odd,
                            \begin{align*}
                                &\sum_{x_1\in\gf_q\setminus \{\frac{(b_1^2+\overline{b_2})\overline{A}^2+(b_1^2+b_2)A^2+(b_2+\overline{b_2})A\overline{A}}{(b_2+\overline{b_2})(A+\overline{A})}\}}(-1)^{\tr\left(\frac{1}{(A+\overline{A})^2}x_1^2+\frac{1}{A+\overline{A}}x_1+1+\frac{A\overline{A}}{(A+\overline{A})^2}+\frac{M_1}{M_0}\right)}+1\\
                                =&\sum_{x_1\in\gf_q\setminus \{\frac{(b_1^2+\overline{b_2})\overline{A}^2+(b_1^2+b_2)A^2+(b_2+\overline{b_2})A\overline{A}}{(b_2+\overline{b_2})(A+\overline{A})}\}}(-1)^{\tr\left(\frac{\sqrt{M_1}}{(b_2+\overline{b_2})(A+\overline{A})x_1+(b_1^2+\overline{b_2})\overline{A}^2+(b_1^2+b_2)A^2+(b_2+\overline{b_2})A\overline{A}}\right)}+1\\
                                =&0.
                            \end{align*}
                \end{itemize}
    \end{itemize}
    Thus, 
    \begin{align*}
        &\sum_{u_1\in U_{q+1}\setminus (R_1\cup R_2)}\frac{1-(-1)^{\tr\left(\frac{[(b_2+\overline{b_2})u_1^2+(b_1^2+b_2)u_1+(b_1^2+b_2)][(b_1^2+\overline{b_2})u_1^2+(b_1^2+\overline{b_2})u_1+b_2+\overline{b_2}]}{[(b_1^2+\overline{b_2})u_1^2+(b_2+\overline{b_2})u_1+b_1^2+b_2]^2}\right)}}{2}\\
        =&\frac{1}{2}\left[\sum_{u_1\in U_{q+1}\setminus (R_1\cup R_2)}1-\sum_{u_1\in U_{q+1}\setminus (R_1\cup R_2)}(-1)^{\tr\left(\frac{[(b_2+\overline{b_2})u_1^2+(b_1^2+b_2)u_1+(b_1^2+b_2)][(b_1^2+\overline{b_2})u_1^2+(b_1^2+\overline{b_2})u_1+b_2+\overline{b_2}]}{[(b_1^2+\overline{b_2})u_1^2+(b_2+\overline{b_2})u_1+b_1^2+b_2]^2}\right)}\right]\\
        =&\begin{cases}
            q, b_2\in\gf_q\setminus\{b_1^2\},\\
            \frac{q-2}{2},b_2\in\gf_{q^2}\setminus\gf_q, s\text{ is odd},\\
            \frac{q-4}{2},b_2\in\gf_{q^2}\setminus\gf_q, s\text{ is even}.\\
        \end{cases}
    \end{align*}
    This implies that when $q>4$, \eqref{eqt:peq2_heq1_1} admits solutions when $b_2\neq b_1^2$. Therefore, when $q>4$, there exists $u_1\in U_{q+1}\setminus(R_1\cup R_2)$ satisfying \eqref{eqt:peq2_heq1_4}. Moreover, if $s$ is odd, we may choose such a $u_1$ with
        $$u_1^2+u_1+1\neq 0.$$
Indeed, at most two elements of $U_{q+1}$ satisfy $u_1^2+u_1+1=0$, while the above counting result shows that the number of admissible choices of $u_1$ is positive even after excluding these two elements. For this choice of $u_1$, let $u_2\in U_{q+1}\setminus\{1,u_1\}$ be a solution of \eqref{eqt:peq2_heq1_3}. Then \eqref{eqt:peq2_heq1_1} holds. Hence the two equations in \eqref{eqtsys:peq2_heq1_a_3} are compatible. We next show that \eqref{eqtsys:peq2_heq1_a_3} determines an element $\alpha_3\in F_q$. Note that the two coefficients of $\alpha_3^2$ in \eqref{eqtsys:peq2_heq1_a_3} cannot vanish simultaneously, otherwise we would have $u_2^2+u_1=0$ and $u_1^2+u_2=0$. This implies $u_1^3=1$, contradicting the choice of $u_1$. Thus at least one equation in \eqref{eqtsys:peq2_heq1_a_3} determines $\alpha_3^2$. Furthermore, it is easy to observe that the $\alpha_3^2$ obtained from \eqref{eqtsys:peq2_heq1_a_3} belongs to $\gf_q$. Then we have $\alpha_3\in \gf_q$ from \eqref{eqtsys:peq2_heq1_a_3} as the map $x\mapsto x^2$ is a bijection on $\gf_q$.
In summary, the solution of \eqref{eqtsys:peq2_heq1_1} could be given by 
    $$\begin{cases}
		    \alpha_1=\frac{u_2^{-1}(b_1+\alpha_3)+u_2(b_1+\alpha_3)}{u_1^{-1}u_2+u_1u_2^{-1}},\\
		    \alpha_2=\frac{u_1^{-1}(b_1+\alpha_3)+u_1(b_1+\alpha_3)}{u_1^{-1}u_2+u_1u_2^{-1}},\\
            u_1\in U_{q+1}\setminus (R_1\cup R_2) \text{ satisfying } \eqref{eqt:peq2_heq1_4} \text{ and } u_1^3+1\neq 0,\\
            u_2\in U_{q+1}\setminus\{u_1,1\} \text{ satisfying } \eqref{eqt:peq2_heq1_3},\\
            u_3=1,\\
            \alpha_3 \text{ satisfying } \eqref{eqtsys:peq2_heq1_a_3}.
	    \end{cases}$$
\end{proof}

Now, we can determine the desired covering radius.
\begin{theorem}\label{thm:qeven_CR}
    When $q$ is even, it holds that
        $$\rho\bigl(\C_{(q,q+1,3,1)}\bigr)=3.$$
\end{theorem}
\begin{proof}
    Recall that $\C_{(q,q+1,3,1)}$ has parameters $[q+1,q-3,d]$ with $d\geqslant 3$. To prove $\rho\bigl(\C_{(q,q+1,3,1)}\bigr)\leqslant 3$, according to Lemma~\ref{lem:CR_PCmatrix}, we just need to prove that, for every $b_1,b_2\in\gf_{q^2}$, there exist $\alpha_i'\in\gf_q$ and $u_i'\in U_{q+1}$ $(i=1,2,3)$ such that the following system holds:
    \begin{equation}\label{eqtsys:even_CR_original}
        \begin{cases}
            \alpha_1'u_1'+\alpha_2'u_2'+\alpha_3'u_3'=b_1,\\
            \alpha_1'(u_1')^2+\alpha_2'(u_2')^2+\alpha_3'(u_3')^2=b_2.
        \end{cases}
    \end{equation}

    Since $q$ is even, every nonzero element of $\gf_{q^2}$ can be uniquely written as $u\beta$, where $u\in U_{q+1}$ and $\beta\in\gf_q^*$. In particular, if $b_2\ne 0$, set $\sqrt{b_2}=u_b\beta_b$, where $u_b\in U_{q+1}$ and $\beta_b\in\gf_q^*$; if $b_2=0$, take $u_b=1$ and $\beta_b=0$. Under the substitutions $\alpha_i'=\alpha_i^2$ and $u_i'=u_bu_i$ $(i=1,2,3)$, system \eqref{eqtsys:even_CR_original} is equivalent to
    \begin{equation}\label{eqtsys:even_CR_normalized}
        \begin{cases}
            \alpha_1u_1+\alpha_2u_2+\alpha_3u_3=\beta_b,\\
            \alpha_1^2u_1+\alpha_2^2u_2+\alpha_3^2u_3=\frac{b_1}{u_b},
        \end{cases}
    \end{equation}
    where $\alpha_i\in\gf_q$ and $u_i\in U_{q+1}$ $(i=1,2,3)$. Here we use the fact that squaring is an automorphism of both $\gf_q$ and $\gf_{q^2}$. By Lemma~\ref{lem:qeven_rho_ubound}, system \eqref{eqtsys:even_CR_normalized} admits a solution for $q>4$. Hence $\rho\bigl(\C_{(q,q+1,3,1)}\bigr)\leqslant 3$ in this case. Together with the matching lower bound from Lemma~\ref{lem:rho_lower_bound}, this proves $\rho\bigl(\C_{(q,q+1,3,1)}\bigr)=3$ for $q>4$.

    For the exceptional case $q=4$, the generator polynomial is
        \[ g_{(4,5,3,1)}(X)=\frac{X^5-1}{X-1}=1+X+X^2+X^3+X^4. \]
    Thus $\C_{(4,5,3,1)}$ is the quaternary repetition code with parameters $[5,1,5]$.
    Every word in $\gf_4^5$ contains a symbol occurring at least twice, so its distance to the corresponding constant codeword is at most three. Therefore $\rho\bigl(\C_{(4,5,3,1)}\bigr)\leqslant 3$. Lemma~\ref{lem:rho_lower_bound} then gives $\rho\bigl(\C_{(4,5,3,1)}\bigr)=3$.
\end{proof}

\subsection{Generalized Covering Radii}
Throughout this subsection, let $H=H_{(q,q+1,3,1)}$. As in Definition~\ref{def:tCR}, for $I\subseteq\{0,\ldots,q\}$, let $H_I$ be the submatrix consisting of the columns of $H$ indexed by $I$, and let $\langle H_I\rangle_{\gf_q}$ denote their $\gf_q$-linear span. 
\begin{theorem}\label{thm:tCR_qeven}
    Let $q=2^m$ with $m\geqslant2$. Then, for every $t\geqslant2$,
        \[ \rho_t(\C)=4. \]
\end{theorem}
\begin{proof}
Lemma~\ref{lem:t_CR_PCmatrix} gives $\rho_t(\C)\leqslant4$ for every integer $t\geqslant1$. Thus it suffices to prove $\rho_2(\C)\geqslant4$. Let $G=\{(a,0):a\in\gf_{q^2}\}$. It is easy to check that $\dim_{\gf_q}G=2$ and $G$ is a $\gf_q$-subspace of $\gf_{q^2}^{2}$. Let $\pi:\gf_{q^2}^{2}\longrightarrow\gf_{q^2}$ be a $\gf_q$-linear projection defined as $\pi(a,b)=b$. We claim that $G\not\subseteq\langle H_I\rangle_{\mathbb F_q}$ whenever $|I|\le3$. First, if $u_1,u_2\in U_{q+1}$ are distinct, then $u_1^2$ and $u_2^2$ are linearly independent over $\gf_q$. Otherwise, a dependence would imply $\frac{u_1^2}{u_2^2}\in\gf_q^{*}\cap U_{q+1}=\{1\}$ and hence $u_1=u_2$, a contradiction. Therefore, if $2\leqslant|I|\leqslant3$, then $\pi(\langle H_I\rangle_{\gf_q})=\gf_{q^2}$.
Applying the rank--nullity theorem to $\pi|_{\langle H_I\rangle_{\gf_q}}$ gives
    \[ \dim_{\gf_q}\bigl(\langle H_I\rangle_{\gf_q}\cap G\bigr)=\dim_{\gf_q}\langle H_I\rangle_{\gf_q}-2\leqslant |I|-2\leqslant1. \]
If $|I|\leqslant1$, then
    \[ \dim_{\gf_q}\bigl(\langle H_I\rangle_{\gf_q}\cap G\bigr)\leqslant\dim_{\gf_q}\langle H_I\rangle_{\gf_q}\leqslant1. \]
Since $\dim_{\gf_q}G=2$, the claim follows.

Fix $b\in\gf_{q^2}\setminus\gf_q$. The syndromes $(1,0)$ and $(b,0)$ form an $\gf_q$-basis of $G$, so they cannot both belong to $\langle H_I\rangle_{\gf_q}$ for any $I\subseteq \{0,\ldots,q\}$ with $|I|\leqslant3$. By the definition of the second generalized covering radius, $\rho_2(\C_{(q,q+1,3,1)})\geqslant4$.
The upper bound and monotonicity now give $\rho_t(\C_{(q,q+1,3,1)})=4$ for every integer $t\geqslant2$.
\end{proof}

\section{Deep Holes}\label{sec:DH} 
In view of Theorem \ref{thm:qeven_CR}, determining the deep holes of \(\mathcal{C}_{(q,q+1,3,1)}\) amounts to identifying the syndromes that cannot be attained by errors of weight at most two. In this section a fairly large number of symbols are employed. Unless otherwise specified, lowercase Greek letters denote elements of \(\mathbb F_q\), and uppercase calligraphic letters denote sets. For $x\in\gf_{q^2}$, $\sqrt{x}$ denotes the unique element satisfying $(\sqrt{x})^2=x$.

We now turn to the syndrome analysis, beginning with the cases in which exactly one component is zero.
\begin{lemma}\label{lem:axes}
      Every syndrome $(a,0)$ or $(0,b)$ with $a\ne0$ and $b\ne0$ is unattainable by an error of weight at most two.
\end{lemma}
\begin{proof}
      If $\lambda_1u_1+\lambda_2u_2=0$ with distinct $u_1,u_2\in U_{q+1}$ and $\lambda_1,\lambda_2\in\gf_q^*$, then $u_1/u_2=\lambda_2/\lambda_1\in U\cap\gf_q^*=\{1\}$, a contradiction. Similarly, $\lambda_1u_1^2+\lambda_2u_2^2=0$ implies $(u_1/u_2)^2=1$, hence $u_1=u_2$.
\end{proof}

It remains to consider syndromes with two nonzero components, for which we first characterize those arising from errors of weight one.
\begin{lemma}\label{lem:syn_weight_1}
      For $ab\ne0$, $(a,b)$ is the syndrome of an error of weight one if and only if
            $$a\bar{a}=b\bar{b}\quad \text{and}\quad a^2\bar{b}+\bar{a}^{2}b=0.$$
\end{lemma}
\begin{proof}
      A weight-one syndrome has the form $(\theta u,\theta u^2)$ with $\theta\in\gf_q^*$ and $u\in U$, and it satisfies $a\bar{a}=b\bar{b}$ and $a^2\bar{b}+\bar{a}^{2}b=0$. Conversely, condition $a^2\bar{b}+\bar{a}^{2}b=(\bar{a}a)^2(ba^{-2}+\bar{b}\bar{a}^{-2})=0$ gives $ba^{-2}\in\gf_q$. Then taking $e=\sqrt{a\bar{a}}$ and $x=a/e\in U$, combining with $a\bar{a}=b\bar{b}$, gives $(a,b)=(ex,ex^2)$.
\end{proof}

Excluding the case covered by Lemma \ref{lem:syn_weight_1}, we now turn to derive a criterion for attainability by an error of weight two.
\begin{lemma}\label{lem:syn_weight_le2}
      Suppose $ab\ne0$, and exclude the weight-one case in Lemma \ref{lem:syn_weight_1}. Then $(a,b)$ is attained by an error of weight at most two if and only if there is $\xi\in\gf_q$ such that
      \begin{equation}\label{eq:cubic}
            \xi^3+b\bar{b}\xi+(a^2\bar{b}+\bar{a}^{2}b)=0,~ \xi^2\ne b\bar{b} ~\text{and}~ \tr\left(\frac{\xi^2+a\bar{a}}{\xi^2+b\bar{b}}\right)=1.
      \end{equation}
      If $a^2\bar{b}+\bar{a}^{2}b=0$, this condition reduces to $\tr(a\bar{a}/b\bar{b})=1$.
\end{lemma}
\begin{proof}
      Suppose first that
      \begin{equation}\label{eq:twoerror}
            a=\lambda_1 u_1+\lambda_2 u_2 ~\text{and}~ b=\lambda_1 u_1^2+\lambda_2 u_2^2,
      \end{equation}
      where $u_1,u_2\in U_{q+1}$, $u_1\ne u_2$ and $\lambda_1,\lambda_2\in\gf_q$. Since $\bar{a}=\lambda_1 u_1^{-1}+\lambda_2 u_2^{-1}$ and $\bar{b}=\lambda_1 u_1^{-2}+\lambda_2 u_2^{-2}$,
      \begin{equation}\label{eq:recurrence}
            a=(u_1+u_2)(\lambda_1+\lambda_2)+(u_1u_2)\bar{a} ~\text{and}~ b=(u_1+u_2)a+(u_1u_2)(\lambda_1+\lambda_2).
      \end{equation}
      If $(\lambda_1+\lambda_2)^2+a\bar{a}=0$, then $\lambda_1+\lambda_2\ne0$, and multiplying the two equations appropriately gives $a(\lambda_1+\lambda_2)+b\bar{a}=(u_1+u_2)((\lambda_1+\lambda_2)^2+a\bar{a})=0$. It follows that $b=a^2/(\lambda_1+\lambda_2)$, a weight-one syndrome by Lemma~\ref{lem:syn_weight_1}, contrary to the assumption. Hence $(\lambda_1+\lambda_2)^2+a\bar{a}\ne0$, and solving \eqref{eq:recurrence} gives
      \begin{equation}\label{eq:sp}
            u_1+u_2=\frac{a(\lambda_1+\lambda_2)+b\bar{a}}{(\lambda_1+\lambda_2)^2+a\bar{a}} ~\text{and}~
            u_1u_2=\frac{a^2+b(\lambda_1+\lambda_2)}{(\lambda_1+\lambda_2)^2+a\bar{a}}.
      \end{equation}
      Substituting \eqref{eq:sp} into $(\overline{u_1+u_2})(u_1u_2)=u_1+u_2$ gives
      \begin{align*}
            &(\bar{a}(\lambda_1+\lambda_2)+b^qa)(a^2+b(\lambda_1+\lambda_2))+(a(\lambda_1+\lambda_2)+b\bar{a})((\lambda_1+\lambda_2)^2+a\bar{a}) \\
            =&a((\lambda_1+\lambda_2)^3+b\bar{b}(\lambda_1+\lambda_2)+a^2\bar{b}+\bar{a}^{2}b)\\
            =&0.
      \end{align*}
      Thus $(\lambda_1+\lambda_2)^3+b\bar{b}(\lambda_1+\lambda_2)+a^2\bar{b}+\bar{a}^{2}b=0$.

      For any $\lambda_1,\lambda_2\in\gf_q$ satisfying $(\lambda_1+\lambda_2)^3+b\bar{b}(\lambda_1+\lambda_2)+a^2\bar b+\bar a^2b=0$, direct expansion yields
      \begin{align}
            \N(a^2+b(\lambda_1+\lambda_2))&=(a\bar{a})^2+(a^2\bar{b}+\bar{a}^{2}b)(\lambda_1+\lambda_2)+b\bar{b}(\lambda_1+\lambda_2)^2 \notag  \\ 
                                          &=((\lambda_1+\lambda_2)^2+a\bar{a})^2,\label{eq:norm1} \\
            \N(a(\lambda_1+\lambda_2)+b\bar{a})&=a\bar{a}(\lambda_1+\lambda_2)^2+(a^2\bar{b}+\bar{a}^{2}b)(\lambda_1+\lambda_2)+a\bar{a}\,b\bar{b} \notag  \\
                                                &=((\lambda_1+\lambda_2)^2+a\bar{a})((\lambda_1+\lambda_2)^2+b\bar{b}).\label{eq:norm2}
      \end{align}
      By $u_1+u_2\ne0$ and \eqref{eq:sp}, \eqref{eq:norm2} implies $(\lambda_1+\lambda_2)^2\ne b\bar{b}$. Moreover, $\overline{(u_1+u_2)}=(u_1+u_2)/(u_1u_2)$, and
      \begin{equation}\label{eq:kappa}
            \frac{u_1u_2}{(u_1+u_2)^2}=\frac1{\overline{(u_1+u_2)}(u_1+u_2)}=\frac{(\lambda_1+\lambda_2)^2+a\bar{a}}{(\lambda_1+\lambda_2)^2+b\bar{b}}\in\gf_q^*.
      \end{equation}
      The locations are the roots of $X^2+(u_1+u_2)X+u_1u_2$, i.e., $Z^2+Z+\frac{u_1u_2}{(u_1+u_2)^2}=0$ after $X=(u_1+u_2)Z$. Its two roots yield distinct elements of $U_{q+1}$ if and only if $\tr(\frac{u_1u_2}{(u_1+u_2)^2})=1$. It indeed holds by Lemma \ref{lem:values_for_addsum} due to $u_1/u_2\in U_{q+1}\setminus\{1\}$. This proves necessity in \eqref{eq:cubic}.

      Conversely, suppose \eqref{eq:cubic} holds, i.e. there exists \(\xi\in\gf_q\) such that
            \[ \xi^3+b\bar{b}\,\xi+(a^2\bar{b}+\bar{a}^{2}b)=0, ~ \xi^2\ne b\bar{b} \text{ and } \tr\left(\frac{\xi^2+a\bar{a}}{\xi^2+b\bar{b}}\right)=1. \]
      If \(\xi^2=a\bar{a}\), the cubic gives
            \[ a^2\bar{b}+\bar{a}^{2}b=\xi(a\bar{a}+b\bar{b}), \]
      and
            \[ \N\left(b+\frac{a^2}{\xi}\right)=b\bar{b}+a\bar{a}+\frac{a^2\bar{b}+\bar{a}^{2}b}{\xi}=0. \]
      This again implies \(b=a^2/\xi\), the excluded weight-one case. Thus \(\xi^2+a\bar{a}\ne0\), and we may define \(u_1+u_2\) and \(u_1u_2\) by \eqref{eq:sp} with \(\xi\) in place of \(\lambda_1+\lambda_2\):
            \[ u_1+u_2=\frac{a \xi+b\bar{a}}{\xi^2+a\bar{a}} \text{ and } u_1u_2=\frac{a^2+b\xi}{\xi^2+a\bar{a}}. \]
      Equations \eqref{eq:norm1}--\eqref{eq:norm2} give \(u_1u_2\ne0\), \(u_1+u_2\ne0\), and \(u_1u_2\in U_{q+1}\). The same expansion used above gives $\overline{(u_1+u_2)}(u_1u_2)=u_1+u_2$. Consequently, (24) holds. By the trace condition, the two roots of $Z^2+Z+\frac{u_1u_2}{(u_1+u_2)^2}=0$ lie in $\gf_{q^2}\setminus\gf_q$ and satisfy $\bar Z=Z+1$. For either root, $\operatorname{N}\bigl((u_1+u_2)Z\bigr)=\frac{(u_1+u_2)^2}{u_1u_2}Z(Z+1)=1.$ Thus $X^2+(u_1+u_2)X+u_1u_2=0$ has two distinct roots $u_1,u_2\in U_{q+1}$.

      Finally, choose the unique \(\lambda_1,\lambda_2\in \gf_{q^2}\) such that
            \[ \lambda_1 u_1+\lambda_2 u_2=a \text{ and } \lambda_1+\lambda_2=\xi. \]
      Then \(\lambda_1+\lambda_2\) is precisely the root \(\xi\) of the cubic equation fixed at the beginning of the converse. Equation \eqref{eq:recurrence} gives
            \[\frac{\lambda_1}{u_1}+\frac{\lambda_2}{u_2}=\frac{a+(u_1+u_2)(\lambda_1+\lambda_2)}{u_1u_2}=\bar{a}.\]
      Taking $q$-th power of the defining equations shows that $(\overline{\lambda_1},\overline{\lambda_2})$ and $(\lambda_1,\lambda_2)$ solve the same nonsingular linear system with coefficient rows $(1,1)$ and $(u_1^{-1},u_2^{-1})$. Thus $\lambda_1,\lambda_2\in\gf_q$. Finally, $\lambda_1 u_1^2+\lambda_2 u_2^2=(u_1+u_2)a+(u_1u_2)(\lambda_1+\lambda_2)=b$, so \eqref{eq:twoerror} holds.

      When $a^2\bar{b}+\bar{a}^{2}b=0$, the cubic is $(\lambda_1+\lambda_2)(\lambda_1+\lambda_2+\sqrt{b\bar{b}})^2=0$. Only $\lambda_1+\lambda_2=0$ satisfies $(\lambda_1+\lambda_2)^2\ne b\bar{b}$, and the trace condition is $\tr(a\bar{a}/(b\bar{b}))=1$.
\end{proof}

Combining Theorem~\ref{thm:qeven_CR} with Lemmas~\ref{lem:axes}--\ref{lem:syn_weight_le2}, we obtain the following characterization of all deep-hole syndromes.
\begin{theorem}\label{thm:even_deep_hole_syndromes}
    Let $q=2^m$ with $m\ge2$. A syndrome $(a,b)\in\mathbb F_{q^2}^2$ is a deep-hole syndrome of $\C_{(q,q+1,3,1)}$ if and only if one of the following holds:
    \begin{enumerate}
        \item Exactly one of $a,b$ is zero.
        \item $ab\ne0$, $a^2\bar b+\bar a^{\,2}b=0$,$a\bar a\ne b\bar b$, and $\tr\!\left(\frac{a\bar a}{b\bar b}\right)=0.$
        \item $ab\ne0$, $a^2\bar b+\bar a^{\,2}b\ne0$, and every $\xi\in\mathbb F_q$ satisfying $\xi^3+b\bar b\,\xi+a^2\bar b+\bar a^{\,2}b=0$ also satisfies
               $\tr\!\left(\frac{\xi^2+a\bar a}{\xi^2+b\bar b} \right)=0.$
    \end{enumerate}
    In the last item, all denominators are nonzero, and the condition holds automatically when the cubic has no root in $\mathbb F_q$.
\end{theorem}
\begin{proof}
By Theorem~\ref{thm:qeven_CR}, the zero syndrome is therefore excluded, and Lemma~\ref{lem:axes} gives the first item.

Suppose $ab\ne0$. If $a^2\bar b+\bar a^{\,2}b=0$, Lemma~\ref{lem:syn_weight_1} excludes precisely the weight-one syndromes by requiring $a\bar a\ne b\bar b$. Lemma~\ref{lem:syn_weight_le2} then shows that unattainability by an error of weight at most two is equivalent to $\operatorname{Tr}(a\bar a/(b\bar b))=0$, giving the second.

If $a^2\bar b+\bar a^{\,2}b\ne0$, Lemma~\ref{lem:syn_weight_1} excludes weight-one syndromes automatically. Moreover, a root $\xi$ of the cubic cannot satisfy $\xi^2=b\bar b$, since substitution would give $a^2\bar b+\bar a^{\,2}b=0$. Since the absolute trace takes values in $\mathbb F_2$, negating the attainability criterion in Lemma~\ref{lem:syn_weight_le2} gives exactly the third.
\end{proof}

It is known that every $a\in \gf_{q^2}^*$ has the unique decomposition $a=\sqrt{a\bar{a}}\cdot u$, where $\sqrt{a\bar{a}}\in\gf_q^*,$ and $u=a/\sqrt{a\bar{a}} \in U_{q+1}$. For $ab\ne0$, define $r=b/(\sqrt{a\bar{a}}u^2)\in\gf_{q^2}^*$. Multiplication of a syndrome by $\sqrt{a\bar{a}}^{-1}$ reduces it to $(u,ru^2)$, where deep-hole condition is invariant under this nonzero multiplication by \eqref{eq:distance}. Obviously, Every $(a,b)$ gives a unique $(u,ru^2)$, and
\begin{align*}
    u\bar{u}=1, ~ (ru^2)^{q+1}=r\bar{r} \text{ and } a^2\bar{b}+\bar{a}^2b=(\sqrt{a\bar{a}})^3(r+\bar{r}).
\end{align*}
If $r\in\gf_q^*$, Theorem~\ref{thm:even_deep_hole_syndromes} gives precisely $r\ne1$ and $\tr(r^{-1})=0$. For $r\notin\gf_q$, set
\begin{equation}\label{eqt:hdelta}
    \beta=(r\bar{r})^{-1/2} \text{ and } \delta=(r+\bar{r})\beta^3.
\end{equation}
To describe $r$ in terms of $\delta$ and $\beta$, we introduce the following quadratic root sets.
\begin{definition}\label{def:set_Q}
    For $\delta\in\gf_q^*$ and $\beta\in\gf_q$ satisfying
    \begin{equation}\label{eq:feasible}
        \tr\left(\frac{\beta}{\sqrt\delta}\right)=1,
    \end{equation}
    define
    \begin{equation}\label{eq:Q}
        \Q(\delta,\beta)=\left\{r\in \gf_{q^2}^*:r^2+\frac{\delta}{\beta^3}r+\frac1{\beta^2}=0\right\}.
    \end{equation}
\end{definition}
\begin{remark}
    Definition \ref{def:set_Q} is valid since condition \eqref{eq:feasible} implies $\beta\ne0$ and guarantees that \eqref{eq:Q} has two distinct conjugate roots in $\gf_{q^2}^*\setminus\gf_q$.
\end{remark}
For $r\in\gf_{q^2}\setminus\gf_q$, let $\beta,\delta$ be defined by \eqref{eqt:hdelta}. The polynomial in \eqref{eq:Q} has roots $r,\bar r$ and is irreducible over $\gf_q$; hence \eqref{eq:feasible} holds and $r\in\mathcal Q(\delta,\beta)$. Conversely, the two roots $r,\bar{r}$ in \eqref{eq:Q} satisfy $r\bar{r}=\beta^{-2}$ and $r+\bar{r}=\delta \beta^{-3}$, so \eqref{eqt:hdelta} recovers the parameters uniquely. Applying Theorem~\ref{thm:even_deep_hole_syndromes} to the normalized syndrome $(u,ru^2)$ and substituting $\xi=z/\beta$, we obtain
\begin{equation}\label{eq:normalizedcubic}
    f_\delta(z)=z^3+z+\delta=0.
\end{equation}
No root is $0$ or $1$ because $\delta\ne0$. The trace condition becomes
\begin{equation}\label{eq:normalizedtrace}
    \tr\left(\frac{\xi^2+1}{\xi^2+r\bar{r}}\right)=\tr\left(\frac{z^2+\beta^2}{z^2+1}\right)=\tr(1)+\tr\left(\frac{\beta+1}{z+1}\right).
\end{equation}
Consequently, the deep-hole condition is equivalent to $\mathrm{Tr}\left(\frac{\beta+1}{z+1}\right)=\mathrm{Tr}(1)$ for every $\gf_q$-root $z$ of $f_\delta$. These reductions motivate the following four sets, corresponding to the case \(r\in\mathbb{F}_q^*\) and, for \(r\notin\mathbb{F}_q\), the cases where \(f_\delta\) has zero, one, or three roots in \(\mathbb{F}_q\).
\begin{definition}\label{def:R}
    Define four subsets of $\gf_{q^2}^*$ as follows.
    \begin{enumerate}
        \item Let
            \begin{equation}\label{eq:R0}
                \R_0=\{\gamma\in\gf_q^* : \gamma\ne1,\ \tr(\gamma^{-1})=0\}.
            \end{equation}
        \item Set
            \[G=\begin{cases}\gf_q^*,&m\text{ even},\\U_{q+1},&m\text{ odd},\end{cases} \qquad \Delta_0=\{v+v^{-1}:v\in G,\ v^{|G|/3}\ne1\}.\]
            Here $3\mid |G|$, and $\Delta_0\subseteq\gf_q^*$. Define
            \begin{equation}\label{eq:R1}
                \R_1=\bigcup_{\delta\in\Delta_0}\ \bigcup_{\substack{\beta\in\gf_q\\\tr(\beta/\sqrt\delta)=1}}\Q(\delta,\beta).
            \end{equation}
        \item For $\zeta\in\gf_q\setminus\{0,1\}$ with $\tr(\zeta^{-1})=\tr(1)+1$, set $\delta_\zeta=\zeta^3+\zeta$ and
            \[ \mathcal{H}_\zeta=\left\{\beta\in\gf_q: \tr\left(\frac{\beta}{\sqrt{\delta_\zeta}}\right)=1,\quad \tr\left(\frac{\beta+1}{\zeta+1}\right)=\tr(1)\right\}. \]
            Define
            \begin{equation}\label{eq:R2}
                \R_2=\bigcup_{\substack{\zeta\in\gf_q\setminus\{0,1\}\\\tr(\zeta^{-1})=\tr(1)+1}} \ \bigcup_{\beta\in \mathcal{H}_\zeta}\Q(\delta_\zeta,\beta).
            \end{equation}
        \item If $m$ is odd, put $\R_3=\varnothing$. If $m$ is even, for $\kappa\in\gf_q\setminus\gf_4$ set
            \[ \zeta_\kappa=\frac{1}{\sqrt{\kappa^2+\kappa+1}},\qquad \delta_\kappa=\frac{\kappa(\kappa+1)}{(\kappa^2+\kappa+1)\sqrt{\kappa^2+\kappa+1}}, \]
            and define
            \[ \mathcal{H}_\kappa=\left\{ \beta\in\gf_q: \begin{aligned}
                                                &\tr(\beta/\sqrt{\delta_\kappa})=1,\\
                                                &\tr((\beta+1)/(\zeta_\kappa+1))=0,\\
                                                &\tr((\beta+1)/(\kappa\zeta_\kappa+1))=0
                                          \end{aligned}\right\}.\]
            Put
            \begin{equation}\label{eq:R3}
                \R_3=\bigcup_{\kappa\in\gf_q\setminus\gf_4}\ \bigcup_{\beta\in \mathcal{H}_\kappa}\Q(\delta_\kappa,\beta).
            \end{equation}
    \end{enumerate}
    Finally, let $\R_q=\R_0\cup\R_1\cup\R_2\cup\R_3$.
\end{definition}
\begin{remark}\label{rem:explicitQ}
    Even the quadratic root operation in \eqref{eq:Q} can be replaced by a fixed linear-algebra construction.  Choose \(e \in \gf_{q^2}\) with \(\bar{e} + e = 1\), and set \(\varepsilon = e^{2} + e \in \mathbb{F}_{q}\); then \(\tr(\varepsilon) = 1\). Fix an \(\mathbb{F}_{2}\)-linear right inverse \(J\) of \(x\mapsto x^{2} + x\) from \(\ker \tr\) to \(\mathbb{F}_{q}\), so that \(J(y)^{2} + J(y) = y\) for \(y\in \ker \tr\). Such a map exists because \(x\mapsto x^{2} + x\) maps \(\mathbb{F}_{q}\) onto \(\ker \tr\). For the parameters in \eqref{eq:feasible}, one has the fully explicit formula
    \begin{equation}\label{eq:Qexplicit}
      \Q(\delta,\beta)=\left\{\frac{\delta}{\beta^3}\left(e+J\left(\frac{\beta^4}{\delta^2}+\varepsilon\right)+i \right) : i\in\gf_2\right\}.
    \end{equation}
    The argument of $J$ has trace zero by $\tr\left(\frac{\beta^4}{\delta^2}\right)=\tr\left(\frac{\beta^2}{\delta}\right)=\tr\left(\frac{\beta}{\sqrt\delta}\right)=1$. Thus Definition~\ref{def:R} together with \eqref{eq:Qexplicit} uses only field arithmetic, a noncube test, and binary linear algebra.
\end{remark}

To analyze the parameter sets in Definition~\ref{def:R}, we first classify the $\gf_q$-roots of $f_\delta$.
\begin{lemma}\label{lem:cubics}
    Let $\delta\in\gf_q^*$. The polynomial $f_\delta$ is separable and has either zero, one, or three roots in $\gf_q$.
    \begin{enumerate}
        \item It has no root in $\gf_q$ if and only if $\delta\in\Delta_0$ from Definition~\ref{def:R}.
        \item It has exactly one root in $\gf_q$ if and only if $\delta=\zeta^3+\zeta$ for some $\zeta\in\gf_q\setminus\{0,1\}$ with $\tr(\zeta^{-1})=\tr(1)+1$. This $\zeta$ is unique.
        \item If it splits over $\gf_q$, its roots can be written $\zeta,\kappa\zeta,(\kappa+1)\zeta$, where
            \[ \kappa\in\gf_q\setminus\{0,1\},\quad \kappa^2+\kappa+1\ne0,\quad \zeta=(\kappa^2+\kappa+1)^{-1/2} \text{ and } \delta=\frac{\kappa(\kappa+1)}{(\kappa^2+\kappa+1)^{3/2}}. \]
            Conversely these formulas always give three distinct roots in $\gf_q$.
    \end{enumerate}
\end{lemma}
\begin{proof}
    The derivative of $f_\delta$ is $(X+1)^2$, and $f_\delta(1)=\delta\ne0$, so $f_\delta$ is separable. The complete proof corresponding to the three cases are given below seperately.
    \begin{enumerate}
        \item The roots of $V^2+\delta V+1$ are $v,v^{-1}$. They either both lie in $\gf_q^*$ or both lie in $U$, and $v\ne1$. Let $H$ be the other group among $\gf_q^*$ and $U_{q+1}$ besides $G$. Since $3\nmid |H|$, $x\mapsto x^3$ is a bijection on $H$. The identity
            \begin{equation}\label{eq:Dickson}
                (w+w^{-1})^3+(w+w^{-1})=w^3+w^{-3}
            \end{equation}
            shows that if $v$ is a cube in either group containing it, then $f_\delta$ has a root $w+w^{-1}$ in $\gf_q$. Conversely, given a root $\zeta\in\gf_q$ of $f_\delta$, choose a root $w$ of $W^2+\zeta W+1$. Then $w\in\gf_q^*$ or $w\in U_{q+1}$, and \eqref{eq:Dickson} gives $w^3\in\{v,v^{-1}\}$. The intersection $\gf_q^*\cap U$ is $\{1\}$. Hence if $v\in G\setminus G^3$, such a $w$ is impossible: it cannot lie in $G$, and if it lay in $H$ then $v$ or $v^{-1}$ would belong to $G\cap H=\{1\}$. Consequently, $f_\delta$ has no root in $\gf_q$ if and only if $v\in G\setminus G^3$. Since $G$ is cyclic and $\delta=v+v^{-1}$, this is equivalent to $\delta\in\Delta_0$.
        \item If $\zeta\in\gf_q$ is a root, then
                \[f_\delta(X)=(X+\zeta)(X^2+\zeta X+\zeta^2+1).\]
                The quadratic factor is irreducible over $\gf_q$ if and only if
                    \[\tr((\zeta^2+1)/\zeta^2)=\tr(1)+\tr(\zeta^{-1})=1.\]
                This proves both the criterion and uniqueness.
        \item The three roots have sum zero. Taking the ratio of an ordered pair of distinct roots gives $\zeta,\kappa\zeta,(\kappa+1)\zeta$ with $\kappa\ne0,1$. By Vieta's theorem, it holds that $\zeta^2(\kappa^2+\kappa+1)=1$ and $\zeta^3\kappa(\kappa+1)=\delta$. The two equations give $\delta=\frac{\kappa(\kappa+1)}{(\kappa^2+\kappa+1)^{3/2}}$. The converse also holds by a direct multiplication of $(x+\zeta)(x+\kappa\zeta)(x+(\kappa+1)\zeta)$.
    \end{enumerate}
\end{proof}

Combining this classification with the trace condition in \eqref{eq:normalizedtrace} yields the following characterization of normalized deep-hole syndromes.
\begin{lemma}\label{lem:paramclassification}
    For any $u\in U$ and $r\in \gf_{q^2}^*$, the syndrome $(u,ru^2)$ is unattainable by an error of weight at most two if and only if $r\in\R_q$. 
\end{lemma}
\begin{proof}
    The case $r\in\gf_q^*$ was settled in $\R_0$ by Theorem \ref{thm:even_deep_hole_syndromes}. For $r\notin\gf_q$, if $f_\delta$ has no $\gf_q$-root, the condition is vacuous, giving $\R_1$ by Lemma~\ref{lem:cubics}(i). If it has exactly one root, Lemmas \ref{lem:cubics} and \eqref{eq:normalizedtrace} give exactly $\R_2$.

    Suppose now that the roots are $\zeta_1,\zeta_2,\zeta_3\in\gf_q$. The logarithmic derivative identity at $X=1$ gives
    \begin{equation}\label{eq:reciprocalroots}
        \sum_{i=1}^3\frac1{\zeta_i+1}=\frac{f_\delta'(1)}{f_\delta(1)}=0.
    \end{equation}
    The required condition at each root, by \eqref{eq:normalizedtrace}, is $\tr((\beta+1)/(\zeta_i+1))=\tr(1)$. Adding the three conditions and using \eqref{eq:reciprocalroots} forces $\tr(1)=0$. Hence no such parameters occur for odd $m$. For even $m$, two of these conditions imply the third. By Lemma~\ref{lem:cubics}, these conditions give exactly $\mathcal R_3$, since the excluded values of $\kappa$ are precisely the four elements of $\gf_4$.
\end{proof}
\begin{corollary}\label{cor:disjoint_qeven}
    The four sets defining $\R_q$ are pairwise disjoint.
\end{corollary}
\begin{proof}
    By Definition~\ref{def:R}, $\mathcal R_0\subseteq\gf_q^*$ and $\mathcal R_i\subseteq\gf_{q^2}\setminus\gf_q$ for $i=1,2,3$. Hence $\mathcal R_0\cap(\mathcal R_1\cup\mathcal R_2\cup\mathcal R_3)=\varnothing$.

    For each $r\in\gf_{q^2}\setminus\gf_q$, equation \eqref{eqt:hdelta} uniquely determines $(\delta,\beta)$. By Definition~\ref{def:R} and Lemma~\ref{lem:cubics}, if $r$ belongs to $\mathcal R_1$, $\mathcal R_2$, or $\mathcal R_3$, then $f_\delta$ has respectively $0$, $1$, or $3$ roots in $\gf_q$. These possibilities are mutually exclusive, so $\mathcal R_1,\mathcal R_2,\mathcal R_3$ are pairwise disjoint.
\end{proof}

Having established the pairwise disjointness of the four parameter sets, we next determine their cardinalities.
\begin{lemma}\label{lem:count}
The parameter sets have the following sizes:
    \[\begin{array}{c|c|c}
        &m\text{ even}&m\text{ odd}\\\hline
        |\R_0| & \frac{q}{2}-2 & \frac{q}{2}-1 \\
        |\R_1| & \frac{q(q-1)}{3} & \frac{q(q+1)}{3} \\
        |\R_2| & \frac{q^2}{4} & \frac{q(q-2)}{4} \\
        |\R_3| & \frac{q(q-4)}{24} & 0.
    \end{array}\]
\end{lemma}
\begin{proof}
    $|\R_0|$ is obvious. For $\R_1$, there are $2|G|/3$ noncubes in $G$ and the pair $\{v,v^{-1}\}$ gives exactly one $v+v^{-1}$. Hence $|\Delta_0|=|G|/3$. Besides, different $(\delta,\beta)$ gives different $\Q(\delta,\beta)$ in Definition \ref{def:set_Q}. Therefore, 
        $$|\R_1|=|\Delta_0|\cdot\frac{q}{2}\cdot2=\frac{q|G|}{3}.$$

    For $\R_2$, similar to $|\R_0|$, we have
        $$ |\{\zeta\in\gf_q\setminus\{0,1\} : \tr(\zeta^{-1})=\tr(1)+1\}|=
            \left\{\begin{aligned}
                &\frac{q}{2}-1,&  & \text{when } m \text{ is odd} \\
                &\frac{q}{2},&  & \text{when } m \text{ is even}
            \end{aligned}\right. $$
    $\frac{1}{\sqrt{\delta_\zeta}}$ and $\frac{1}{\zeta+1}$ are $\gf_2$-linearly independent otherwise $\delta_\zeta=(\zeta+1)^2$ and $\delta_\zeta=\zeta^3+\zeta$ lead to $\zeta=1$. Therefore, $\beta \mapsto \tr(\beta/\sqrt{\delta_\zeta})$ and $\beta \mapsto \tr(\beta/(\zeta+1))$ are also linearly independent over \(\mathbb F_2\). This implies
        $$|\mathcal{H}_\zeta|=\frac{q}{4}.$$
    Therefore,
        $$|\R_2|=2\cdot|\{\zeta\in\gf_q\setminus\{0,1\} : \tr(\zeta^{-1})=\tr(1)+1\}|\cdot|\mathcal{H}_\zeta|=\left\{\begin{aligned}
                &\frac{q}{2}(\frac{q}{2}-1), & & \text{when } m \text{ is odd} \\
                &\frac{q^2}{4}, & & \text{when } m \text{ is even}
            \end{aligned}\right.$$     
    
    For $\mathcal R_3$, the case of odd $m$ is immediate since $\mathcal R_3=\varnothing$, so it remains to consider even $m$. Note that different choices of $\kappa$ may give rise to the same polynomial $f_\delta$ that splits completely over $\gf_q$. Hence, we first consider the number of completely split polynomials \(f_\delta\) over \(\mathbb F_q\), denoting it by \(n_{f_\delta}\). For a fixed such \(f_\delta\), it has three roots in \(\mathbb F_q\) and the ratio of any two distinct roots is of the form \(\kappa\). Because there are \(3\cdot 2=6\) ordered pairs of distinct roots, the same cubic polynomial is associated with six distinct values of \(\kappa\). The resulting six ratios are
        \[ \kappa,\quad \kappa^{-1},\quad \kappa+1,\quad (\kappa+1)^{-1},\quad \kappa/(\kappa+1),\quad (\kappa+1)/\kappa. \]
    They are distinct when $\kappa\notin\gf_4$: a fixed point of a nonidentity transformation in this list would satisfy $\kappa\in\{0,1\}$ or $\kappa^2+\kappa+1=0$. Thus $$n_{f_\delta}=\frac{q-4}{6}.$$ 
    For a given $\kappa$, 
            $$|\mathcal{H}_\kappa|=\sum\limits_{\beta\in\gf_q}\frac{1-(-1)^{\tr\left(\frac{\beta}{\sqrt{\delta_\kappa}}\right)}}{2}\frac{1+(-1)^{\tr\left(\frac{\beta+1}{\zeta_\kappa+1}\right)}}{2}\frac{1+(-1)^{\tr\left(\frac{\beta+1}{\kappa\zeta_\kappa+1}\right)}}{2}=\frac{q}{8}.$$
    In short, $$|\R_3|=2\cdot n_{f_\delta}|\mathcal{H}_\kappa|=\frac{q(q-4)}{24}.$$
\end{proof}

Combining the preceding classification and counting results, we obtain the following explicit description of all deep holes of $\C$, together with nonredundant representatives and enumeration formulas.
\begin{theorem}\label{thm:deepholes_even}
Let $q=2^m\ge4$. Let $\R_q$ be given by Definition~\ref{def:R}, and let $\ee(a,b)$ be given by \eqref{eq:lagrange-basis}--\eqref{eq:representative}. Define
\begin{equation}\label{eq:D}
    \mathcal{D}=\{\ee(u,0):u\in U\} \cup\{\ee(0,u):u\in U\} \cup\{\ee(u,ru^2):u\in U,\ r\in\R_q\}.
\end{equation}
Then the following assertions hold.
    \begin{enumerate}
    \item All deep holes of $\C_{(q,q+1,3,1)}$ are given by the disjoint union
            \begin{equation}\label{eq:DH}
                \mathrm{D}\mathrm{H}(\C_{(q,q+1,3,1)})=\bigsqcup_{\mathbf d\in \mathcal{D}}\ \bigsqcup_{\lambda\in\gf_q^\times}(\lambda\mathbf d+\C_{(q,q+1,3,1)}).
            \end{equation}
            In particular, $\mathcal{D}$ is a complete and nonredundant set of representatives under equivalence, and
            \[ \{\lambda\mathbf d:\lambda\in\gf_q^\times,\ \mathbf d\in \mathcal{D}\} \] is a complete and nonredundant set of representatives of all deep-hole cosets. Every displayed representative is supported on coordinates $0,1,2,3$.
    \item The four sets in Definition~\ref{def:R} are pairwise disjoint, and
            \begin{equation}\label{eq:Rcount}
                |\R_q|=\begin{cases}
                    \frac{5q^2}{8}-2,&m\text{ even},\\
                    \frac{7q^2+4q-12}{12},&m\text{ odd}.
                \end{cases}
                \end{equation}
    \item Put \begin{equation}\label{eq:M}
                M_q=|\R_q|+2=\begin{cases}
                        \frac{5q^2}{8},&m\text{ even},\\
                        \frac{7q^2+4q+12}{12},&m\text{ odd}.
                        \end{cases}
                \end{equation}
            The numbers of deep-hole cosets and deep-hole vectors are $(q^2-1)M_q$ and $q^{q-3}(q^2-1)M_q$, respectively.
    \end{enumerate}
\end{theorem}
\begin{proof}
    By \eqref{eq:distance} and $\rho(\C_{(q,q+1,3,1)})=3$, deep-hole syndromes are precisely those unattainable by errors of weight at most two. Since $q$ is even, every nonzero element of $\gf_{q^2}$ factors uniquely as $\alpha u$, with $\alpha\in\gf_q^*$ and $u\in U_{q+1}$. Thus every nonzero syndrome has a unique expression of one of the forms $\alpha(u,0)$, $\alpha(0,u)$, $\alpha(u,ru^2),$ where $r\in\gf_{q^2}^*$ in the third form. By Lemmas~\ref{lem:axes} and~\ref{lem:paramclassification}, the first two forms always give deep-hole syndromes, whereas the third does so exactly when $r\in\R_q$.

    Lemma~\ref{lem:interpolation} therefore identifies all deep-hole syndromes with the nonzero multiples of the syndromes of the representatives in \eqref{eq:D}. The uniqueness above ensures that these multiples are distinct. Since syndrome fibers are exactly code cosets, \eqref{eq:DH} is a disjoint union of all deep holes, and both representative assertions follow. The support assertion follows from \eqref{eq:representative}.

    Corollary~\ref{cor:disjoint_qeven} and Lemma~\ref{lem:count} give the second assertion. With $M_q=|\R_q|+2$, since each coset contains $q^{q-3}$ vectors, the numbers of deep-hole cosets and deep-hole vectors are $(q^2-1)M_q$ and $q^{q-3}(q^2-1)M_q,$ respectively, as stated in last item.
\end{proof}

\section{Coset leader weight distribution}\label{sec:CLWD}
Having determined the covering radius in Theorem \ref{thm:qeven_CR} and enumerated the deep-hole cosets in Theorem \ref{thm:deepholes_even}, we now derive the complete coset-leader weight distribution of \(\mathcal{C}_{(q,q+1,3,1)}\).

\begin{theorem}\label{thm:coset-leader-distribution_even}
    Let $q=2^m\geq4$. The complete coset-leader weight distribution is
    \begin{equation} \label{eq:bch-coset-leader-distribution_even}
        \begin{array}{c|c|c}
            i & m\ \text{even} & m\ \text{odd} \\
            \hline
            0 & 1 & 1 \\[2pt]
            1 & q^2-1 & q^2-1 \\[2pt]
            2 & \displaystyle\frac{3q^2(q^2-1)}{8} & \displaystyle\frac{(q^2-1)(5q^2-4q-12)}{12} \\[6pt]
            3 & \displaystyle\frac{5q^2(q^2-1)}{8} & \displaystyle\frac{(q^2-1)(7q^2+4q+12)}{12}
        \end{array}
    \end{equation}
    and $N_i(\C_{(q,q+1,3,1)})=0$ for $i>3$. Equivalently, with $M_q=|\mathcal{R}_q|+2$ as in Theorem~\ref{thm:deepholes_even}, the coset-leader weight enumerator is
    \begin{equation} \label{eq:coset-leader-enumerator_even}
        1+(q^2-1)Z+(q^2-1)(q^2-M_q)Z^2+(q^2-1)M_qZ^3.
    \end{equation}
\end{theorem}
\begin{proof}
    It is clear that there are exactly $q^4$ cosets. The zero syndrome corresponds to $\C_{(q,q+1,3,1)}$ itself, giving $N_0(\C_{(q,q+1,3,1)})=1$.

    A weight-one error with nonzero entry $\lambda\in\mathbb{F}_q^*$ at the coordinate indexed by $x\in U_{q+1}$ has syndrome $(\lambda x,\lambda x^2)$. These syndromes are nonzero and pairwise distinct. Consequently,
        \[ N_1(\C_{(q,q+1,3,1)})=|U_{q+1}|(q-1)=(q+1)(q-1)=q^2-1. \]

    Since $\rho(\C_{(q,q+1,3,1)})=3$, a coset has coset-leader weight $3$ if and only if it is a deep-hole coset. Theorem~\ref{thm:deepholes_even} therefore gives
        \[ N_3(\C_{(q,q+1,3,1)})=(q^2-1)M_q. \]
    The four possible coset-leader weights partition all cosets, whence
        \[ N_2(\C_{(q,q+1,3,1)})=q^4-N_0(\C_{(q,q+1,3,1)})-N_1(\C_{(q,q+1,3,1)})-N_3(\C_{(q,q+1,3,1)})=(q^2-1)(q^2-M_q). \]
    Substituting the two values of $M_q$ from Theorem~\ref{thm:deepholes_even} proves the table and the enumerator.
\end{proof}

The following two examples illustrate the closed-form evaluation for even $q$ in Theorem \ref{thm:coset-leader-distribution_even} and have been checked by direct enumeration with a MAGMA program.
\begin{example}
When $q=2^3$, formula \eqref{eq:M} gives $|\mathcal R_q|=39$ and $M_q=41$. By Theorem \ref{thm:coset-leader-distribution_even}, the coset-leader weight distribution of $C_{(q,q+1,3,1)}$ is
    \[ N_0=1,\quad N_1=63,\quad N_2=1449,\quad N_3=2583, \]
equivalently, its coset-leader weight enumerator is
    \[ 1+63Z+1449Z^2+2583Z^3, \]
which coincides with the result directly computed by a MAGMA program.
\end{example}

\begin{example}
When $q=2^4$, formula \eqref{eq:M} gives $|\mathcal R_q|=158$ and $M_q=160$. By Theorem \ref{thm:coset-leader-distribution_even}, the coset-leader weight distribution of $C_{(q,q+1,3,1)}$ is
    \[ N_0=1,\quad N_1=255,\quad N_2=24480,\quad N_3=40800, \]
equivalently, its coset-leader weight enumerator is
    \[ 1+255Z+24480Z^2+40800Z^3, \]
which coincides with the result directly computed by a MAGMA program.
\end{example}

\section{conclusion}\label{sec:con}
In this paper, we determined the generalized covering radii, the deep holes, and the complete coset-leader weight distribution of \(\mathcal{C}_{(q,q+1,3,1)}\) for \(q=2^m\ge4\). The covering radius is \(3\), whereas every generalized covering radius of order at least two is \(4\). More precisely, 
\begin{itemize}
        \item Theorem~\ref{thm:qeven_CR} establishes $\rho(\C_{(q,q+1,3,1)})=3$ and Theorem~\ref{thm:tCR_qeven} establishes $\rho_t(\C_{(q,q+1,3,1)})=4$ for $t\geqslant2$, mainly by calculating character sums.
        \item Theorem~\ref{thm:even_deep_hole_syndromes} gives necessary and sufficient conditions for deep-hole syndromes through analyzing the roots of cubic equations.
        \item Theorem~\ref{thm:deepholes_even} explicitly describes all deep holes and provides complete nonredundant representatives of their cosets. The construction combines explicit parametrization and Lagrange interpolation.
        \item Theorem~\ref{thm:coset-leader-distribution_even} determines the complete coset-leader weight distribution.
    \end{itemize}
Unlike the even-characteristic case studied in this paper, the odd-characteristic case requires a different approach and will be pursued in future work. More broadly, extending the generalized covering-radius hierarchy, deep-hole classification, and coset-leader weight distribution to other families of antiprimitive BCH codes and to general linear codes remains an interesting direction for our future work.


\begin{thebibliography}{99}
    \bibitem{BlokhuisPS2022} A. Blokhuis, R. Pellikaan, T. Szőnyi, ``The extended coset leader weight enumerator of a twisted cubic code,'' \emph{Des. Codes Cryptogr.}, vol. 90, pp. 2223-2247, 2022.
    \bibitem{Charpin1994} P. Charpin, ``Weight distributions of cosets of two-error-correcting binary BCH codes, extended or not,'' {\it IEEE Trans. Inform. Theory},vol. 40, no. 5, pp. 1425--1442, 1994.
    \bibitem{CharpinHellesethZinoviev2006} P. Charpin, T. Helleseth, and V. A. Zinoviev, ``The coset distribution of triple-error-correcting binary primitive BCH codes,'' {\it IEEE Trans. Inform. Theory}, vol. 52, no. 4, pp. 1727--1732, 2006.
    \bibitem{Cohen1997} S. D. Cohen, ``The length of primitive BCH codes with minimal covering radius,'' \emph{Des. Codes Cryptogr.}, vol. 10, no. 1, pp. 5--16, 1997.
    \bibitem{Ding2016} C. Ding, ``BCH codes in the past 55 years'', in {\it Proc. 7th Int. Workshop Finite Fields Appl.}, Tianjin, China, 2016.
    \bibitem{DingLi2024} C. Ding and C. Li, ``BCH cyclic codes,'' \emph{Discrete Math.}, vol. 347, no. 5, Art. 113918, 2024.
    \bibitem{DingTang2020} C. Ding and C. Tang, ``Infinite families of near MDS codes holding $t$-designs,'' {\it IEEE Trans. Inform. Theory}, vol. 66, no. 9, pp. 5419--5428, Sep. 2020.
    \bibitem{tCRIT21} D. Elimelech, M. Firer and M. Schwartz, ``The generalized covering radii of linear codes'', {\it IEEE Trans. Inform. Theory}, vol. 67, no. 12, pp. 8070--8085, 2021.
    \bibitem{FangWW2018} W.~Fang, L.-P.~Wang, and D.~Wan, ``On deep holes of Gabidulin codes,'' \emph{Finite Fields Appl.}, vol.~54, pp.~216--229, 2018.
    \bibitem{FangXZ2025} W. Fang, J. Xu and R. Zhu, ``Deep holes of twisted {R}eed-{S}olomon codes'', {\it Finite Fields Appl.}, vol. 108, Paper No. 102680, 31, 2025.
    \bibitem{GorensteinEtAl1960} D. Gorenstein, W. W. Peterson, and N. Zierler, ``Two-error correcting Bose--Chaudhuri codes are quasi-perfect,'' \emph{Inf. Control}, vol. 3, no. 3, pp. 291--294, 1960.
    
    \bibitem{HaradaO2004} M. Harada, M. Ozeki and K. Tanabe, On the covering radius of ternary extremal self-dual codes, {\it Des. Codes Cryptogr.} vol. 33, no.~2, pp. 149--158, 2004.
    \bibitem{Helleseth1978} T. Helleseth, ``All binary 3-error-correcting BCH codes of length $2^m-1$ have covering radius 5 (Corresp.)'', {\it IEEE Trans. Inform. Theory}, vol. 24, no. 2, pp. 257-258, 1978.
    \bibitem{Helleseth1979}T. Helleseth, ``The weight distribution of the coset leaders for some classes of codes with related parity-check matrices'', {\it Discrete Math.}, vol. 28, no. 2, pp. 161--171, 1979.
    \bibitem{Hou1994} X.-D. Hou, ``Classification of cosets of the Reed--Muller code $R(m-3,m)$,'' \emph{Discrete Math.}, vol. 128, nos. 1--3, pp. 203--224, 1994.
    \bibitem{HuffmanPless2003} W. C. Huffman and V.~S. Pless, Fundamentals of error-correcting codes, Cambridge Univ. Press, Cambridge, 2003.
    \bibitem{Kaipa2017} K. Kaipa, ``Deep holes and MDS extensions of Reed-Solomon codes'', {\it IEEE Trans. Inform. Theory}, vol. 63, no. 8, pp. 4940-4948, 2017.
    \bibitem{KavutTutdere2019} S. Kavut and S. Tutdere, ``The covering radii of a class of binary cyclic codes and some BCH codes,'' \emph{Des. Codes Cryptogr.}, vol. 87, nos. 2--3, pp. 317--325, 2019.
    \bibitem{Klapper2006} A.~Klapper and A.~Mertz, ``The two covering radius of the two error correcting BCH code,'' in \emph{Proc. IEEE Int. Symp. Inf. Theory (ISIT)}, Seattle, WA, USA, 2006, pp.~2497--2501.

    \bibitem{LiDingLi2017} C. Li, C. Ding, and S. Li, ``LCD cyclic codes over finite fields,'' {\it IEEE Trans. Inform. Theory}, vol. 63, no. 7, pp. 4344--4356, July 2017.
    \bibitem{LiuEtAl2022} Q. Liu, C. Ding, S. Mesnager, C. Tang, and V. D. Tonchev, ``On Infinite Families of Narrow-Sense Antiprimitive BCH Codes Admitting 3-Transitive Automorphism Groups and Their Consequences,'' {\it IEEE Trans. Inform. Theory}, vol. 68, no. 5, pp. 3096-3107, 2022.
    \bibitem{LiuDingLi2017} H. Liu, C. Ding, and C. Li, ``Dimensions of three types of BCH codes over GF(q)'', {\it Discrete Math.}, vol. 340, no. 8, pp. 1910-1927, 2017.
    \bibitem{McLoughlin1984} A. McLoughlin, ``The complexity of computing the covering radius of a code,'' {\it IEEE Trans. Inform. Theory}, vol. 30, pp. 800-804, 1984.
    
    \bibitem{OrozcoJ2024} J.~C. Orozco and H. Janwa, Resolution of a conjecture on the covering radius of linear codes, in {\it Combinatorics, graph theory and computing}, 413--425, Springer Proc. Math. Stat., 448, Springer, Cham, 2024.
    \bibitem{Ozbudak2026} F.~\"Ozbudak and I.~\"Ozt\"urk, ``The third generalized covering radius for binary primitive double-error-correcting BCH codes,'' \emph{Finite Fields Appl.}, vol.~110, Paper No.~102749, 2026.
    \bibitem{ShiTO2022} M. Shi, T. Helleseth, F. Özbudak and P. Solé, ``Covering Radius of Melas Codes'', {\it IEEE Trans. Inform. Theory}, vol. 68, no. 7, pp. 4354-4364, 2022.
    \bibitem{ShiHO2023} M. Shi, T. Helleseth and F. \"Ozbudak, ``Covering radius of generalized Zetterberg type codes over finite fields of odd characteristic'', {\it IEEE Trans. Inform. Theory}, vol. 69, no. 11, pp. 7025-7048, 2023.
    \bibitem{ShiLHO2025} M. Shi, S. Li, T. Helleseth and F. Özbudak, "Determining the Covering Radius of All Generalized Zetterberg Codes in Odd Characteristic," {\it IEEE Trans. Inform. Theory}, vol. 71, no. 5, pp. 3602-3613, May 2025.
    \bibitem{TanEtAl2023} P. Tan, C. Fan, C. Ding, C. Tang, and Z. Zhou, ``The minimum locality of linear codes,'' \emph{Des. Codes Cryptogr.}, vol. 91, no. 1, pp. 83--114, Jan. 2023.
    \bibitem{Tzr2018} Z. Tu, X. Zeng, C. Li and T. Helleseth, ``A class of new permutation trinomials'', {\it Finite Fields Appl.}, vol. 50, pp. 178-195, 2018.
    \bibitem{Wei1991} V. K. Wei, ``Generalized Hamming weights for linear codes,'' {\it IEEE Trans. Inform. Theory}, vol. 37, no. 5, pp. 1412-1418, 1991.
    \bibitem{XiongCHY2026} M. Xiong, C. H. Yip, ``On generalized covering radii of binary primitive double-error-correcting BCH codes'', arXiv:2603.21068, 2026.
    \bibitem{XiongY2026} M. Xiong and H. Yan, ``On covering radius of generalized Zetterberg codes'', {\it IEEE Trans. Inform. Theory}, doi: 10.1109/TIT.2026.3716210,2026.
    \bibitem{XuEtAl2025a} H. Xu, X. Wu, W. Lu, and X. Cao, ``The sufficient and necessary conditions for the minimum distance of the BCH code $\mathcal{C}_{(q,q+1,3,h)}$ to be 3 and 4,'' {\it IEEE Trans. Inform. Theory}, vol. 71, no. 6, pp. 4206--4213, June 2025.
    \bibitem{Yohananov2025} L.~Yohananov and M.~Schwartz, ``The second generalized covering radius of binary primitive double-error-correcting BCH codes,'' \emph{Finite Fields Appl.}, vol.~107, Paper No.~102648, 2025.
    \bibitem{EzaharMM1987} M. H. El-Zahar and M. K. Khairat, ``On the weight distribution of the coset leaders of the first-order {R}eed-{M}uller code'', {\it IEEE Trans. Inform. Theory}, vol. 33, no. 5, pp. 744--747, 1987.
    \bibitem{ZhangWK2020} J. Zhang, D. Wan and K. Kaipa, ``Deep holes of projective Reed-Solomon codes,'' {\it IEEE Trans. Inform. Theory}, vol. 66, no. 4, pp. 2392-2401, 2020.
    \bibitem{ZhangW2023} J. Zhang and D. Wan, ``On deep holes of elliptic curve codes'', {\it IEEE Trans. Inform. Theory}, vol. 69, no. 7, pp. 4498--4506, 2023.
    \bibitem{ZhuEtAl2022} H. Zhu, M. Shi, X. Wang, and T. Helleseth, ``The \(q\)-Ary Antiprimitive BCH Codes,'' {\it IEEE Trans. Inform. Theory}, vol. 68, no. 3, pp. 1683-1695, 2022.
    
\end{thebibliography}
\end{document}